\documentclass[pra,reprint,groupedaddress]{revtex4-1}

\pdfoutput=1

 \usepackage{amsmath}
 \usepackage{amsfonts}
 \usepackage{graphicx}
 \usepackage{newtxtext,newtxmath}
\usepackage{dcolumn}

 \usepackage{hyperref}
 \hypersetup{
    colorlinks=true,       
    linkcolor=blue,          
    citecolor=blue,        
    filecolor=magenta,      
    urlcolor=blue           
}

\newcommand{\al}[1]{\begin{align}#1\end{align}}

\newcommand{\Zeff}{Z_\text{eff}}
\renewcommand{\vec}[1]{\boldsymbol{\mathbf{#1}}}

\renewcommand{\epsilon}{\varepsilon}
\newcommand{\eb}{\varepsilon_b}

\newcommand{\pd}[2]{\frac{\partial #1}{\partial #2}}
\newcommand{\pdd}[3]{\frac{\partial^2 #1}{\partial #2 \, \partial #3}}
\newcommand{\pds}[2]{\frac{\partial^2 #1}{\partial #2^2}}

\allowdisplaybreaks

\newcolumntype{d}[1]{D{.}{.}{#1}}

\let\originalleft\left
\let\originalright\right
\renewcommand{\left}{\mathopen{}\mathclose\bgroup\originalleft}
\renewcommand{\right}{\aftergroup\egroup\originalright}

\begin{document}

\frenchspacing

\title{Model-potential calculations of positron binding, scattering, and annihilation for atoms and small molecules, using a Gaussian basis}
\author{A. R. Swann}
\email{a.swann@qub.ac.uk}
\author{G. F. Gribakin}
\email{g.gribakin@qub.ac.uk}
\affiliation{
School of Mathematics and Physics, Queen's University Belfast, University Road, Belfast BT7 1NN, United Kingdom}
\date{\today}

\begin{abstract}
A model-potential method is employed to calculate binding, elastic scattering, and annihilation of positrons for a number of atoms and small nonpolar molecules, namely, Be, Mg, He, Ar, H$_2$, N$_2$, Cl$_2$, and CH$_4$. The model potential contains one free parameter for each type of atom within the target. Its values are chosen to reproduce existing \textit{ab initio} positron-atom binding energies or scattering phase shifts. The calculations are performed using a Gaussian basis for the positron states, and we show how to obtain values of the scattering phase shifts and normalized annihilation rate $\Zeff$ from discrete positive-energy pseudostates.
Good agreement between the present results and existing calculations and experimental data, where available, is obtained, including the $\Zeff $ value for CH$_4$, which is strongly enhanced by a low-lying virtual positron state. An exception is the room-temperature value of $\Zeff$ for Cl$_2$, for which the present value is much smaller than the experimental value obtained over 50 years ago. Our calculations predict that among the molecular targets studied, only Cl$_2$ might support a bound state for the positron, with a  binding energy of a few meV.
\end{abstract}

\maketitle

\section{Introduction}

We have recently proposed a model-potential approach that enables one to calculate the energies and annihilation rates for positron bound states with molecules, including large alkanes \cite{Swann18,Swann19}. In this paper we show that the method can also be used to describe low-energy positron scattering and annihilation in small nonpolar molecules. We also validate it by performing binding, scattering, and annihilation calculations for a number of atoms for which accurate theoretical predictions are available.

The positron ($e^+$) is an important tool in many areas of science, e.g., in tests of QED and the standard model \cite{Karshenboim05,Ishida14,ALEPH06}, astrophysics \cite{Guessoum14}, condensed-matter physics \cite{Tuomisto13}, and in medical imaging \cite{Wahl02}. However, the basic interactions of positrons with ordinary matter are still not fully understood. In particular, this concerns the problem of low-energy positron annihilation in molecules and its resonant enhancement, and the related problem of positron binding to neutral atoms and molecules.

The ability of certain neutral atoms to support bound states for positrons was suggested by many-body-theory calculations in 1995 \cite{Dzuba95} and rigorously proved by variational calculations of the $e^+$Li binding energy two years later \cite{Ryzhikh97,Strasburger98}. A plethora of calculations for other atoms followed (see Ref.~\cite{Mitroy02} for a 2002 review), and it is now expected that about 50 atoms in their ground states can bind a positron \cite{Harabati14}.
Unfortunately, positron-atom bound states have not yet been observed experimentally, though several detection schemes have been proposed \cite{Mitroy99,Dzuba10,Surko12,Swann16}.

Conversely, positron binding energies have been determined experimentally for about 85 polyatomic molecules \cite{Barnes03,Barnes06,Young07,Young08,Young08a,Danielson09,Danielson10,Danielson12,Natisin:thesis,private19}. This has been done by making use of \textit{resonant annihilation}. When a positron collides with a molecule, it can annihilate with a target electron ``in flight.'' Additionally, for polyatomic molecules, annihilation can also proceed by capture of the positron into a bound state, its excess energy being transferred into excitation of a vibrational mode with near-resonant energy  \cite{Gribakin00,GribakinNewDirections,Gribakin10}. This results in  resonances in the annihilation rate at positron energies $\epsilon_\nu = \hbar\omega_\nu - \eb $, where $\eb $ is the positron binding energy and $\omega_\nu$ is the vibrational frequency of mode $\nu$. The binding energy is thus measured as a downshift of the resonance energy with respect to that of the vibrational mode.

Note that resonant annihilation can occur only for molecules that support a bound state for the positron \cite{Surko88,Gilbert02}. The vast majority of molecules studied experimentally to date are nonpolar or weakly polar, e.g., alkanes, arenes, alcohols, formates, and acetates. On the side of theory, calculations of positron-molecule binding have had limited success. Most studies have considered strongly polar molecules, i.e., those with a dipole moment greater than the critical value of 1.625~D that guarantees binding even at the static level of theory \cite{Fermi47,Crawford67}. (For molecules that are free to rotate, the critical dipole moment is greater, and it increases with the molecule's angular momentum \cite{Garrett71}.) In fact, only six species have been studied both theoretically and experimentally, namely, carbon disulfide CS$_2$, acetaldehyde C$_2$H$_4$O, propanal C$_2$H$_5$CHO, acetone (CH$_3$)$_2$CO, acetonitrile CH$_3$CN, and propionitrile C$_2$H$_5$CN \cite{Koyanagi13,Tachikawa12,Tachikawa03,Tachikawa11,Tachikawa14}.
The best agreement is currently at the level of 25\% for acetonitrile, where a configuration-interaction calculation gave $\eb=135$~meV \cite{Tachikawa11}, compared to the measured binding energy of 180~meV \cite{Danielson10}. On the other hand, the calculation found no binding for CS$_2$ \cite{Koyanagi13}, while the experiment gives $\eb=75$~meV \cite{Danielson10}. The calculations are difficult because of strong electron-positron correlation effects that are hard to describe in a complete manner \textit{ab initio}. An overview of calculations of positron-molecule binding carried out to date can be found in Ref.~\cite{Swann18}.

Recently, we proposed a model-potential method for calculating positron-molecule binding energies. In this method, the electrostatic potential of the molecule is first calculated at the static (Hartree-Fock) level. The Schr\"odinger equation is then solved for a positron  in this potential, with the addition of a model potential that accounts for the long-range polarization of the molecule and short-range correlations \cite{Swann18}.
We tested this idea by examining positron binding to hydrogen cyanide HCN \cite{Swann18} and obtained good agreement with existing \textit{ab initio} calculations \cite{Chojnacki06,Kita09}. However, the true strength of our approach is that it can be easily applied to large systems. In Ref. \cite{Swann19} we used it to study positron binding to alkanes with up to 16 carbon atoms. We found good agreement between the calculated and measured binding energies, and we also computed the rates of positron annihilation from the bound states.

In our method \cite{Swann18,Swann19}, the positron wave function is expanded in a basis of square-integrable Gaussian functions. Here we show that in spite of the absence of true continuum, the method can be adapted to calculate low-energy positron scattering and direct annihilation for nonpolar molecules. To test the idea, we first perform calculation for a number of atoms, both positron-binding (Be and Mg) and nonbinding (He and Ar), where accurate calculations exist. Our model positron-molecule correlation potential contains just one free parameter (viz., the cutoff radius) for an atomic target, or one free parameter for each type of atom within a molecular target. Their values for Be, Mg, He, Ar, and H are taken from existing model-potential calculations of positron binding, scattering, and annihilation with atoms \cite{Mitroy02a}, or adjusted to reproduce many-body-theory scattering phase shifts \cite{Green14}. Our calculations for molecular targets, viz., H$_2$, N$_2$, Cl$_2$, and CH$_4$, are more predictive in nature. Here, we calculate $s$-wave scattering phase shifts, scattering lengths, and annihilation rates for all species. We also explore the possibility of positron binding to Cl$_2$ and make comparisons with existing theoretical and experimental data.

Atomic units (a.u.) are used throughout. 

\section{Theory and numerical implementation}

\subsection{Schr\"odinger equation for positron}

The details of our model-potential treatment of the positron-molecule interaction are given in Ref.~\cite{Swann18}. Here we briefly repeat the salient features for convenience.

The nonrelativistic Hamiltonian for a positron interacting with an atomic or molecular target with $N_e$ electrons and $N_a$ nuclei (treated in the Born-Oppenheimer approximation) is
\al{
H = \sum_{i=1}^{N_e} h^e(\vec{r}_i) + h^p(\vec{r}) + \sum_{i=1}^{N_e} \sum_{j<i} \frac{1}{\lvert \vec{r}_i - \vec{r}_j \rvert} - \sum_{i=1}^{N_e} \frac{1}{\lvert \vec{r} - \vec{r}_i \rvert},
}
where 
\al{
h^e(\vec{r}_i) &= -\frac12 \nabla_i^2 - \sum_{A=1}^{N_a} \frac{Z_A}{\lvert \vec{r}_i - \vec{r}_A\rvert} , \\
h^p(\vec{r})&= -\frac12 \nabla^2 + \sum_{A=1}^{N_a} \frac{Z_A}{\lvert \vec{r} - \vec{r}_A\rvert},
}
$\vec{r}_i$ is the position of electron $i$, $\vec{r}_A$ is the position of nucleus $A$ (with charge $Z_A$), and $\vec{r}$ is the position of the positron, all relative to an arbitrary origin. 
A direct solution of the Schr\"odinger equation $H\Psi=E\Psi$ for the total energy $E$ and the $(N_e+1)$-particle wave function $\Psi(\vec{r}_1,\dotsc,\vec{r}_{N_e},\vec{r})$ is numerically intractable for systems with more than a few electrons. 
We therefore proceed by first calculating the energy and wave function of the \textit{bare} target (i.e., without the positron) in its ground state, using the Hartree-Fock method. This wave function $\Phi(\vec{r}_1,\vec{r}_2,\dotsc,\vec{r}_{N_e})$ is a Slater determinant of the $N_e$ electronic spin orbitals.
The positron-target interaction is then taken to be
\al{\label{eq:tot_pot}
V(\vec{r}) = V_\text{st}(\vec{r}) + V_\text{cor}(\vec{r}),
}
where $V_\text{st}$ is the electrostatic potential of the target, calculated at the Hartree-Fock level, and $V_\text{cor}$ accounts for the correlation effects beyond the frozen-target Hartree-Fock approximation. 

In what follows, we assume that the target is closed-shell; thence there are $N_e/2$ doubly occupied electronic molecular orbitals $\varphi_i$, and the electrostatic potential of the target is given by
\al{
V_\text{st}(\vec{r}) = \sum_{A=1}^{N_a} \frac{Z_A}{\lvert \vec{r}-\vec{r}_A\rvert} -2\sum_{i=1}^{N_e/2} \int \frac{\lvert \varphi_i(\vec{r}')\rvert^2}{\lvert \vec{r}-\vec{r}'\rvert} \, d\tau',
}
where $d\tau '$ is the  volume element associated with $\vec{r}'$. The correlation potential $V_\text{cor}$ can be derived using  many-body theory  \cite{Amusia76,Dzuba93,Dzuba95,Dzuba96,Gribakin04,Green14,Green18}, or approximated
using density-functional-theory approaches, based on the positron correlation energy in an electron gas and correct long-range asymptotic form \cite{Jain83,Jain91}. It can also be represented  by a model potential with correct long-range behavior and parametrized form at short range. This approach has long been used for studying low-energy electron-molecule scattering (see, e.g., Ref.~\cite{Burke72}). Model potentials have been used previously to study positron interactions with atoms and polar molecules (see, e.g., Refs.~\cite{Mitroy02a,Mitroy02,Gribakin15,Sugiura18}).

So far, the many-body-theory approach has only been developed for atoms \footnote{The main difficulty here is to provide an accurate description of the important virtual-positronium contribution to $V_\text{cor}$. This effect also presents a major challenge for standard quantum-chemistry approaches, making for very slow convergence with respect to the size of the electron and positron basis sets.}. The density-functional-theory-based approach has been used in quantum-chemistry calculations of positron-molecule binding \cite{Sugiura19} but lacks the quantitative accuracy. Following Refs.~\cite{Swann18,Swann19}, we use a model potential, viz.,
\al{\label{eq:pol_pot}
V_\text{cor}(\vec{r}) = -\sum_{A=1}^{N_a} \frac{\alpha_A}{2\lvert \vec{r}-\vec{r}_A\rvert^4} \left[ 1-\exp\left( -\frac{\lvert \vec{r}-\vec{r}_A\rvert^6}{\rho_A^6}\right)\right],
}
where $\alpha_A$ is the \textit{hybrid} dipole polarizability of atom $A$ within the target \cite{Miller90} (which for an atomic target is just the usual atomic dipole polarizability), and $\rho_A$ is a cutoff radius specific to atom $A$. 
At large distances from the target, $V_\text{cor}(\vec{r})\simeq -\alpha/2r^4$ (where $\alpha=\sum_A\alpha_A$ is the total polarizability of the target and $r$ is the distance from the target to the positron), which is the asymptotic polarization potential for a positron interacting with a closed-shell atom or spherical-top molecule. For molecules with anisotropic polarizabilities, the asymptotic form of $V_\text{cor}(\vec{r})$ is the spherical average of the anisotropic polarization potential \footnote{If the target is not a closed-shell atom or spherical-top molecule, the polarizability tensor $\alpha _{ij}$ is not isotropic, and the true long-range behavior of the positron-molecule interaction potential is $V(\vec{r})\simeq -\frac12 \sum _{ij}\alpha _{ij}E_iE_j={-}\frac12 r^{-6}\sum_{i,j}\alpha_{ij}x_i x_j $, where $E_i=-x_i/r^3$ are the components of the positron electric field at the molecule, and $x_i$ ($i=1$, 2, 3) are the positron Cartesian coordinates with respect to the molecule.}. The function in brackets in Eq.~(\ref{eq:pol_pot}) moderates the unphysical growth of the potential near nucleus $A$. Values of $\rho_A$ correlate with the radius of  atom $A$ and are typically in the range 1.5--3.0~a.u. \cite{Mitroy02a,Swann18,Swann19}. 

The short-range part of $V_\text{cor}$ effectively parametrizes  correlation effects other than polarization, such as virtual positronium (Ps) formation. This latter is notoriously difficult to describe in an \textit{ab initio} manner, as it requires large numbers of electron and positron partial waves \cite{Bray93,Gribakin04}, or addition of basis states centered at points outside the molecule \cite{Varella02,Barbosa17}. In our approach, the precise analytical form of the short-range cutoff function will not strongly affect the results, so long as the cutoff radii $\rho_A$ are chosen judiciously.

The single-particle Schr\"odinger equation for the  positron, 
\al{\label{eq:positron_schrodinger}
\left[-\frac12 \nabla^2 + V(\vec{r}) \right] \psi(\vec{r}) = \epsilon\psi(\vec{r}),
}
is solved to obtain the positron energy $\epsilon$ and wave function $\psi(\vec{r})$,
for the total potential (\ref{eq:tot_pot}). This is referred to as the frozen-target-plus-polarization (FT+P) method in Ref.~\cite{Swann18}. The total wave function of the positron-target system is the product of the electronic Slater determinant and the positron wave function:
\al{\label{eq:Psi}
\Psi(\vec{r}_1,\vec{r}_2,\dotsc,\vec{r}_{N_e},\vec{r}) = \Phi(\vec{r}_1,\vec{r}_2,\dotsc,\vec{r}_{N_e}) \psi(\vec{r}).
}

In practice, Eq.~(\ref{eq:positron_schrodinger}) is solved by expanding $\psi (\vec{r})$ in a Gaussian basis (see below). If the potential $V(\vec{r})$ is sufficiently attractive, the eigenvalue spectrum will contain negative energies, which correspond to the positron bound states. The binding energy $\eb$ for such a state is related to its negative energy eigenvalue $\epsilon$ by $\eb=\lvert\epsilon\rvert$. However, most (or all, for nonbinding species) of the energy eigenvalues $\epsilon $ are positive, and the corresponding wave functions represent positron pseudostates that span the continuum. As shown in Secs. \ref{subsec:elscat} and \ref{subsec:zeff}, they can be used to obtain information on positron scattering and direct annihilation by the target.

\subsection{Numerical details}

The Hartree-Fock electronic molecular orbitals and the resulting electrostatic potential of the molecule are calculated using \textsc{gamess} \cite{Schmidt93,Gordon05}. The Schr\"odinger equation for the positron, Eq.~(\ref{eq:positron_schrodinger}), is solved using the \textsc{neo} package \cite{Webb02,Adamson08}, which we have modified to include the model correlation potential $V_\text{cor}$ in the Roothaan equation for the positron \cite{Swann18}.

The electron and positron wave functions are written in terms of Gaussian basis sets, with several Gaussian primitives centered on each of the atomic nuclei, viz.,
\al{
\varphi_i(\vec{r}_i) &= \sum_{A=1}^{N_a} \sum_{j=1}^{N_A^e} C_{Aj}^{(i)} g_{Aj}(\vec{r}_i) , \\
\psi(\vec{r}) &= \sum_{A=1}^{N_a} \sum_{j=1}^{N_A^p} C_{Aj}^{(p)} g_{Aj}(\vec{r}) , \label{eq:pos_wfn_exp}
}
where 
\al{\label{eq:gaussian_primitive}
g_{Aj}(\vec{r}) =  (x-x_A)^{n_{Aj}^x} (y-y_A)^{n_{Aj}^y} (z-z_A)^{n_{Aj}^z} e^{-\zeta_{Aj} \lvert \vec{r}-\vec{r}_A\rvert^2}
}
is a Gaussian primitive with  angular momentum $n_{Aj}^x + n_{Aj}^y + n_{Aj}^z$. There are $N_A^e$ ($N_A^p$) Gaussian primitives centered on nucleus $A$ for the electron (positron).

For the electrons, the standard 6--311++G($d$,$p$) basis set has been used throughout. 
The geometry of each molecule (assumed to be in its rovibrational ground state) is optimized at the Hartree-Fock level, using this basis.
For the positron, an even-tempered Gaussian basis has been adopted. For the functions of a specific angular-momentum type (e.g., $s$, $p$, $d$) centered on nucleus $A$, we choose
the exponents $\zeta_{Aj}$ as
\al{
\zeta_{Aj} = \zeta_{A1} \beta^{j-1} \qquad (j=1,\dotsc,N_A^p),
}
where $\zeta_{A1}>0$ and $\beta>1$ are parameters. In principle, different choices of $\zeta_{A1}$, $\beta$, and $N_A^p$ can be made for each nucleus.
A prudent choice of the smallest exponents $\zeta_{A1}$ is very important for an accurate description of weakly bound positron states. At large distances, the bound-state wave function behaves as $\psi(\vec{r})\sim e^{-\kappa r}/r$, where $\kappa=\sqrt{2\epsilon_b}$ and $\epsilon_b$ is the binding energy. To ensure that expansion (\ref{eq:pos_wfn_exp}) accurately describes the wave function at distances $r\sim 1/\kappa$, we must have $\zeta_{A1} \lesssim \kappa^2=2\eb$. For nonbinding targets, the value of $\zeta_{A1}$ determines the energy of the lowest positive-energy pseudostate, $\epsilon \sim \zeta_{A1}$.

For the atomic targets (Be, Mg, He, and Ar), we have used two different positron basis sets. The first consists of 12 $s$-type Gaussians with $\zeta_{A1}=0.0001$ and $\beta=3$. The second consists of 19 $s$-type Gaussians with $\zeta_{A1}=0.0001$ and $\beta=2$. These shall be referred to as the $12s$ and $19s$ basis sets, respectively. 

For H$_2$, we have used three different basis sets. The first is identical to the $12s$ basis set used for the atomic targets, but with 12 $s$-type Gaussians centered on \textit{each} H atom (making a total of 24 basis functions). The second set is obtained by taking the first set and adding eight $p$-type Gaussians on each H atom (each of these Gaussians has three projections, making a total of $2\times8\times3=48$ additional basis functions), the values of the exponents $\zeta_{Aj}$ starting from 0.0081 and increasing with a common ratio of $\beta =3$. The third set is obtained by taking the second set and adding eight $d$-type Gaussians on each H atom (each of these Gaussians has six projections, making a total of $2\times8\times6=96$ additional basis functions), again with $\zeta_{A1}=0.0081$ and $\beta =3$. These shall be referred to as the $12s$, $12s\,8p$, and $12s\,8p\,8d$ basis sets, respectively. 

For N$_2$ and Cl$_2$, we use the $12s\,8p\,8d$ basis set again. Finally, for CH$_4$, we use the $12s\,8p\,8d$ basis functions on the C atom, and on each of the H atoms, we use a set of 8 $s$-type Gaussians, with $\zeta_{A1}=0.0081$ and $\beta =3$. This shall be referred to as the $12s\,8p\,8d\,/\,8s$ basis set.

In Ref. \cite{Mitroy02a}, Mitroy and Ivanov used $V_\text{cor}$ in the form of Eq.~(\ref{eq:pol_pot}) (with a single term in the sum over $A$) to investigate positron interactions with a number of atomic targets, including Be, Mg, He, Ar, and H. They determined appropriate values of the cutoff radius $\rho_A$ for each atom by comparing the model-potential calculations with accurate \textit{ab initio} calculations. For Be and Mg, for which a positron bound state exists, they chose $\rho_A$ so that the binding energy $\eb$ fit a stochastic-variational calculation \cite{Mitroy01}. For He and Ar, which do not bind the positron, they determined $\rho_A$ by comparing the $s$-wave scattering phase shift at the positron momentum $k=0.1$~a.u. with the Kohn-variational calculation \cite{vanReeth99} (He) or polarized-orbital calculation \cite{McEachran79} (Ar). For H, they determined $\rho_A$ by comparing the scattering length with close-coupling calculations \cite{Mitroy95,Mitroy95a}.

For Be, Mg, He, and Ar, we use the same values of $\alpha_A$ and $\rho_A$ as given in Table I of Ref.~\cite{Mitroy02a}. This enables a direct comparison of our results for the binding energies, scattering phase shifts, and annihilation rates with those of Ref.~\cite{Mitroy02a}. For Ar, we also carry out calculations for a cutoff radius of $\rho_\text{Ar}=1.88$~a.u., chosen to reproduce the $s$-wave scattering phase shift from the many-body-theory calculations by Green \textit{et al.} \cite{Green14}. For H$_2$, N$_2$, CH$_4$, and Cl$_2$, we use the atomic hybrid polarizabilities from Ref.~\cite{Miller90}. We take the cutoff radius for H to be $\rho_\text{H}=2.051$~a.u. \cite{Mitroy02a}. For C and N, we take the cutoff radius to be the same as for H (as was done in Refs.~\cite{Swann18,Swann19}). Finally, for Cl we use either $\rho_\text{Cl}=1.88$~a.u., i.e., the second cutoff radius of Ar (its periodic-table neighbor), or $\rho_\text{Cl}=2.20$~a.u., chosen to reproduce the experimental binding energy $\eb =57$~meV for CCl$_4$ \cite{Natisin:thesis}. The latter value of the cutoff radius is in accord with the fact that the mean radius of the valence orbital in Cl is 10\%  larger than that of Ar \cite{Radtsig86}. These data are summarized in Table \ref{tab:polarizabilities_cutoffs}.
\begin{table}
\caption{\label{tab:polarizabilities_cutoffs}Values of atomic (He, Be, Mg, Ar) or
hybrid (H, C, N, Cl) polarizabilities $\alpha_A$, and cutoff radii $\rho_A$ used in the correlation potential, Eq.~(\ref{eq:pol_pot}).}
\begin{ruledtabular}
\begin{tabular}{ccc}
Atom $A$ & $\alpha_A$ (a.u.) & $\rho_A$ (a.u.) \\
\hline
He & 1.383 & 1.500 \\
Be & 38 & 2.686 \\
Mg & 72 & 3.032 \\
Ar & 11.1 & 1.710, 1.88\\
H & 2.612 & 2.051 \\
C & 7.160 & 2.051\\
N & 6.451 & 2.051 \\
Cl & 15.62 & 1.88, 2.20
\end{tabular}
\end{ruledtabular}
\end{table}

\subsection{Elastic scattering}\label{subsec:elscat}

After solving the Schr\"odinger equation (\ref{eq:positron_schrodinger}), the positive-energy pseudostates can be used to extract information about positron elastic scattering from the target. Specifically, we can find values of the $s$-wave scattering phase shift for a set of discrete energies \footnote{The scattering by a nonpolar molecule with a center of symmetry can be described as $s$-wave scattering, as long as the mixing between the positron $s$ and $d$ waves can be neglected, i.e., for $kR_a\ll 1$, where $R_a$ is the radius of the target.}. We restrict our interest to the low-energy region $\epsilon < 0.5$~a.u., which corresponds to positron momenta $k<1$~a.u.

For a spherically symmetric (atomic) target, each positive-energy pseudostate has a definite orbital angular momentum $l$. The wave function of each pseudostate factorizes into radial and angular parts as
\al{\label{eq:spherical_separation}
\psi(\vec{r}) = \frac{1}{r} P(r) Y_{lm}(\Omega) ,
}
where $Y_{lm}$ is a spherical harmonic, with $m$ the magnetic quantum number. We restrict our interest to $s$-wave scattering, and ignore the pseudostates with $l>0$. For atoms, our positron basis sets contain only $s$-type Gaussians, so all of the pseudostates do, in fact, have $l=0$, with $Y_{00}=1/\sqrt{4\pi }$.

For molecular targets, the lack of spherical symmetry means that the pseudostates do not have a definite angular momentum. However, for small nonpolar molecules, the mixing of the positron partial waves by the potential is small at low positron energies \cite{Note3}. Hence, we can select the ``$s$-type'' pseudostates, for which the expectation value of the squared orbital angular momentum operator $L^2$ (see Appendix \ref{sec:L2expec}) is close to zero.

For a true positron continuum state with $l=0$, the asymptotic form of the radial wave function is
\al{\label{eq:swave_asymp}
P(r) \simeq \frac{A \sin(kr + \delta_0)}{k},
}
where $A$ is a normalization constant and $\delta_0(k)$ is the $s$-wave scattering phase shift. One way to find $\delta_0$, is to fit the radial function $P(r)$ for a positive-energy pseudostate to the asymptotic form (\ref{eq:swave_asymp}) with $k=\sqrt{2\epsilon }$ at intermediate values of $r$, for which the potential $V(\vec{r})$ is negligible compared to the positron energy, while the wave function is still described well by the Gaussian basis. For $s$-type states in molecules, we can also spherically average the positron wave function around the molecular center of mass (see Appendix \ref{sec:spherical_averaging}), before fitting to Eq.~(\ref{eq:swave_asymp}).

However, we found that fitting the pseudostate wave functions to Eq.~(\ref{eq:swave_asymp}) resulted in values of $\delta_0$ that were sensitive to the range of $r$ used for the fit, making it difficult to obtain reliable phase shifts in this way. We have therefore adopted an alternative and more universal method which allowed us to determine the phase shifts using only the energy eigenvalues. In this method one first solves the Schr\"odinger equation for a \textit{free} positron [i.e., Eq.~(\ref{eq:positron_schrodinger}) with $V(\vec{r})=0$] using a Gaussian basis. This gives a discrete set of positive-energy pseudostates, and as before, we retain only the $s$-type states.  We denote the energies of these states by $\epsilon_n^{(0)}$, where $n=1,2,\dotsc$. Since these energies increase monotonically with $n$, there exists an invertible function $f$ of a continuous variable $n$ such that
\al{
f(n) = \epsilon_n^{(0)}
}
for positive integer $n$.

Solving the Schr\"odinger equation for the positron in the field of the target [i.e., Eq.~(\ref{eq:positron_schrodinger}) with $V(\vec{r})$ given by Eq.~(\ref{eq:tot_pot})] in the same basis, and retaining only the $s$-type states, yields a different set of energy eigenvalues, which we denote $\epsilon_n$. If the positron-target potential supports one or more bound $s$-type levels, the corresponding value(s) of $\epsilon_n$ will be negative. 
Let $\Delta \epsilon_n$ denote the difference between $\epsilon_n$ and the corresponding free-particle energy $\epsilon_n^{(0)}$, viz., $\Delta\epsilon_n=\epsilon_n-\epsilon_n^{(0)}$. A positive (negative) value of $\Delta\epsilon_n$ indicates that the positron-target interaction is effectively repulsive (attractive) at the positron energy $\epsilon_n$, and consequently one expects the $s$-wave scattering phase shift to be smaller (greater) than $N_s\pi$, where $N_s$ is the number of bound $s$ levels supported by the potential \footnote{Recall Levinson's theorem \cite{LandauQM} which relates the value of the phase shift $\delta _0(k)$ at zero momentum to the number of bound states, $\delta _0(0)=N_s\pi $, given that $\delta _0(k)\to 0$ for $k\to \infty $.}. For $n>N_s$, i.e., for the positive-energy pseudostates, the energies $\epsilon_n$ are related to the $s$-wave phase shift by
\al{\label{eq:en_delta0}
\epsilon_n = f \left( n - \frac{\delta_0}{\pi} \right).
}
Equation (\ref{eq:en_delta0}) is inverted to determine the phase shift as
\al{
\delta_0 = \left[n - f^{-1} \left( \epsilon_n\right) \right] \pi ,
}
where the functional inverse $f^{-1}$ can be obtained by plotting integer $n$ against $\epsilon_n^{(0)}$ and constructing a continuous function $f^{-1}(\epsilon )$ by interpolation. 

In practice, for even-tempered Gaussian basis sets, the energies $\epsilon_n^{(0)}$ and $\epsilon_n$ grow approximately exponentially with $n$, making $f(n)$ a rapidly changing function. Hence, we plot values of $n$ against $\ln\epsilon_n^{(0)}$ and interpolate them to obtain a function $g(\ln \epsilon)\equiv f^{-1}( \epsilon)$. Since the values of $\ln\epsilon_n^{(0)}$ grow in a near-linear fashion with $n$, this interpolation is accurate and robust. The phase shift at the positron energy $\epsilon_n$ is then given by
\al{\label{eq:delta0}
\delta_0 = \left[ n - g\left( \ln\epsilon_n\right)\right] \pi .
}

\begin{figure}
\includegraphics[width=3.375in]{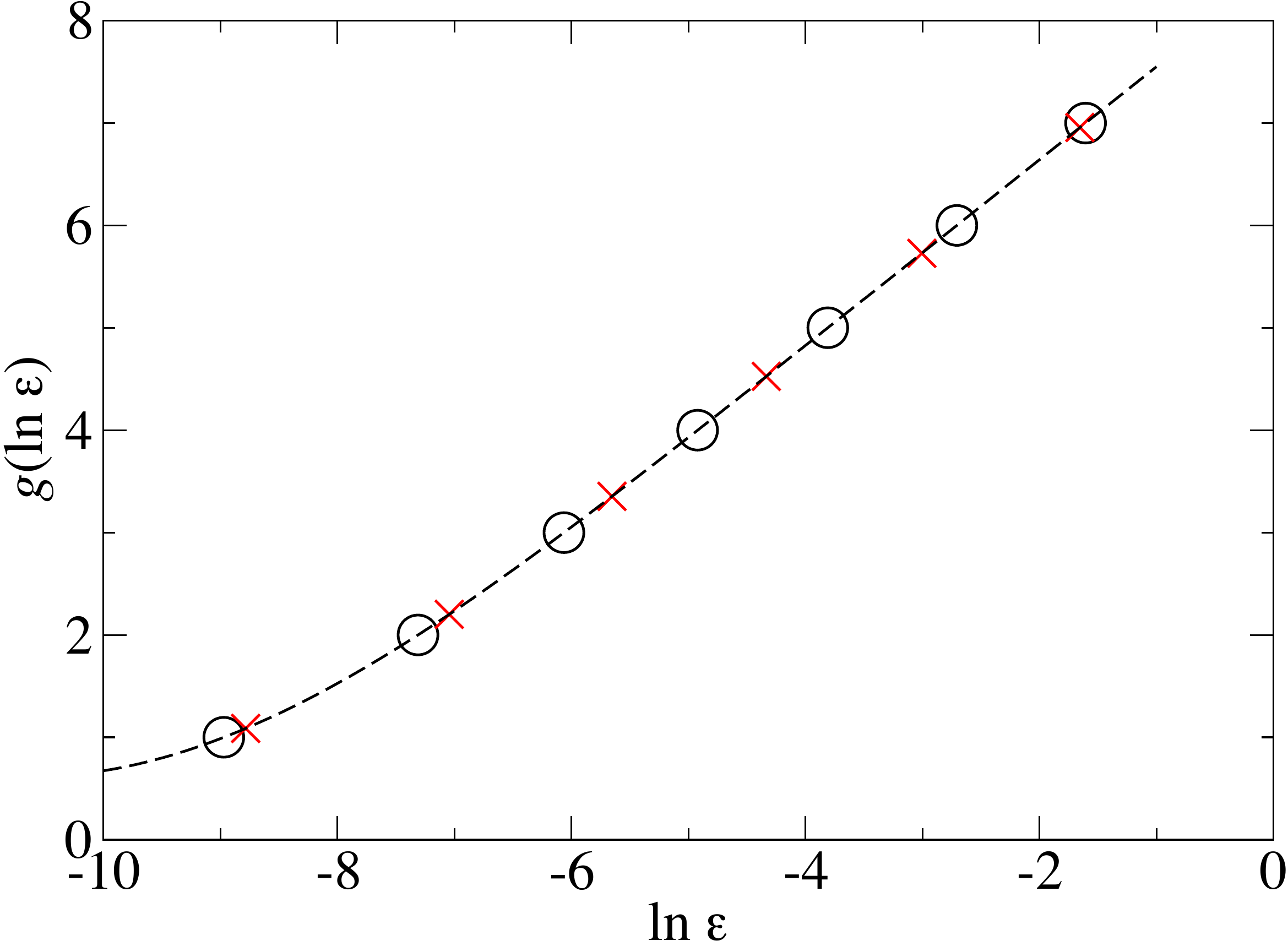}
\caption{\label{fig:Be_fn_12s}The function $g(\ln\epsilon)$ for the $12s$ atomic basis set. Black circles correspond to the integer values $n=g(\ln\epsilon_n^{(0)})$ for
$n=1$--7; dashed lines is a cubic-spline fit of the above data; red crosses correspond to the energies $\epsilon_n$ obtained for a positron interacting with the Be atom.}
\end{figure}

As an example, in Fig.~\ref{fig:Be_fn_12s}, black circles show $n$ plotted against $\ln\epsilon_n^{(0)}$, where the $\epsilon_n^{(0)}$ are the free-particle energies computed using the $12s$ atomic basis set [recall that $g(\ln\epsilon)=n$ for $\epsilon=\epsilon_n^{(0)}$]. Only the data for the seven states with $\epsilon_n^{(0)}<0.5$~a.u. are shown. The black dashed line is the function $g(\ln\epsilon)$ obtained as a cubic-spline fit to the free-particle data. Finally, the figure shows the values of $\ln\epsilon_n$ and the corresponding values of $g(\ln\epsilon_n)$ for a positron in the field $V(\vec{r})$ of the Be atom (red crosses). Note that Be has a bound $s$ state for the positron ($\epsilon _1<0$), so the first cross corresponds to $n=2$. The phase shifts obtained in this way are shown in Sec.~\ref{sec:results}.

Considering the phase shift $\delta_0$ as a function of the positron momentum $k=\sqrt{2\epsilon}$, and fitting to one or more terms of the effective-range-theory expansion \cite{Spruch60}
\al{\label{eq:mert}
k\cot\delta_0 = -\frac{1}{a} + \frac{\pi \alpha }{3a^2}k + O\left(k^2\ln Ck\right)
}
at small $k$, provides estimates of the scattering length $a$. If the positron-target potential supports a weakly bound $s$ state with binding energy $\epsilon_b=-\epsilon _1>0$, the scattering length will be positive and large in magnitude, and related to the binding energy by $\eb \simeq 1/2a^2$. In contrast, a large negative scattering length indicates the presence of a low-lying virtual $s$ level with energy $\epsilon\approx 1/2a^2$. In either case, the zero-energy elastic scattering cross section $\sigma=4\pi a^2$ is much greater than the geometrical cross-sectional area of the target \cite{LandauQM}. The rate of positron direct annihilation is similarly enhanced \cite{Dzuba96,Gribakin00}, e.g., as observed in Ar, Kr, and Xe \cite{Green14} (see Secs. \ref{subsec:zeff} and \ref{sec:results}).

\begin{figure}
\centering
\includegraphics[width=3.375in]{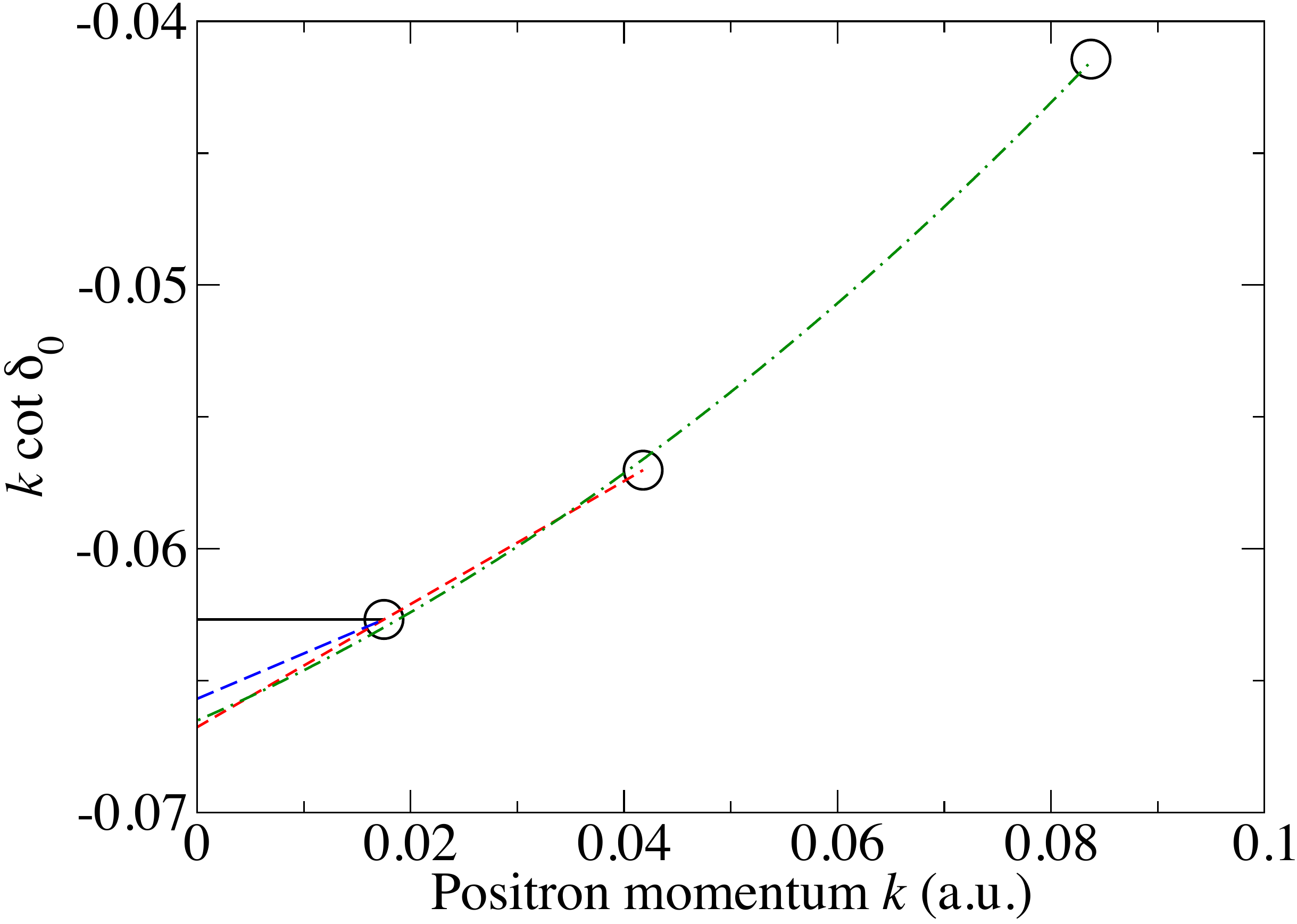}
\caption{\label{fig:phase_example_Be}Values of $k\cot\delta_0$ for Be for the lowest three positive-energy pseudostates, as calculated in the $12s$ basis set. Black circles, calculated values; solid black line, fit (\ref{eq:Be_scatlength_fit_a}); short-dashed red line, fit (\ref{eq:Be_scatlength_fit_b}); long-dashed blue line, fit (\ref{eq:Be_scatlength_fit_c}); dot-dashed green line, fit (\ref{eq:Be_scatlength_fit_d}).}
\end{figure}

As an example, Fig.~\ref{fig:phase_example_Be} shows the values of $k\cot\delta_0$ for Be for the lowest three positive-energy pseudostates, as calculated in the $12s$ basis set.
Also shown are four different fits, based on Eq.~(\ref{eq:mert}), that have been used to estimate the scattering length $a$, as follows:
\begin{subequations}
\al{
k\cot\delta_0 &= -\frac1a , \label{eq:Be_scatlength_fit_a}\\
k\cot\delta_0 &= -\frac1a + Ck , \label{eq:Be_scatlength_fit_b}\\
k\cot\delta_0 &= -\frac1a + \frac{\pi \alpha }{3a^2}k , \label{eq:Be_scatlength_fit_c}\\
k\cot\delta_0 &= -\frac1a + \frac{\pi \alpha }{3a^2}k + C_1 k^2 \ln C_2k , \label{eq:Be_scatlength_fit_d}
}
\end{subequations}
where $a$ (the scattering length), $C$, $C_1$, and $C_2$ are fitting parameters, and  $\alpha=38$~a.u. is the polarizability of Be. Fits (\ref{eq:Be_scatlength_fit_a}) and (\ref{eq:Be_scatlength_fit_c}) use only the lowest-momentum value of $\delta _0$; fit (\ref{eq:Be_scatlength_fit_b})  uses only the first two pseudostates, and fit (\ref{eq:Be_scatlength_fit_d}) uses all three pseudostates. The resulting estimates of the scattering length are $a=15.95$ [fit (\ref{eq:Be_scatlength_fit_a})], 14.98 [fit (\ref{eq:Be_scatlength_fit_b})], 15.22 [fit (\ref{eq:Be_scatlength_fit_c})], and 15.03~a.u. [fit (\ref{eq:Be_scatlength_fit_d})]. The two estimates that are closest to each other are those obtained using fits (\ref{eq:Be_scatlength_fit_b}) and (\ref{eq:Be_scatlength_fit_d}). Consequently, we will use the simpler of these fits, Eq.~(\ref{eq:Be_scatlength_fit_b}), throughout, i.e., perform a linear fit for $k\cot\delta_0$ in terms of $k$ using the two lowest-energy pseudostates.

\subsection{Annihilation from a bound state}

For targets that support a bound state for the positron, we can evaluate the corresponding $2\gamma$ annihilation rate,
\al{\label{eq:Gamma}
\Gamma = \pi r_0^2 c \delta_{ep}.
}
Here $r_0$ is the classical electron radius, $c$ is the speed of light,  $\delta_{ep}$ is the average electron density at the positron,
\al{\label{eq:delta_ep}
\delta_{ep} = \int \sum_{i=1}^{N_e} \delta(\vec{r}-\vec{r}_i) \left\lvert\Psi(\vec{r}_1,\dotsc,\vec{r}_{N_e},\vec{r})\right\rvert^2 \, d\tau_1 \dotsm d\tau_{N_e}  d\tau,
}
also known as the contact density, and $\Psi(\vec{r}_1,\dotsc,\vec{r}_{N_e},\vec{r})$ is the total wave function of the positron bound state, normalized to unity.
The lifetime of the positron bound state is  $1/\Gamma$.

In the independent-particle approximation, the total wave function has the form of Eq.~(\ref{eq:Psi}), and Eq.~(\ref{eq:delta_ep}) reduces to
\al{\label{eq:deltaep_ipa}
\delta_{ep}
= 2\sum_{i=1}^{N_e/2}  \int \left\lvert \varphi_i(\vec{r}) \right\rvert^2  \left\lvert \psi(\vec{r}) \right\rvert^2 \, d\tau,
}
with the electron and positron wave functions normalized as
\al{
\int  \left\lvert \varphi_i(\vec{r}) \right\rvert^2 \, d\tau &= 1 , \\
\int  \left\lvert \psi(\vec{r}) \right\rvert^2 \, d\tau &= 1. \label{eq:posnorm}
}
The independent-particle approximation does not account for the electron-positron Coulomb attraction at short range that increases the contact density. As a result, Eq.~(\ref{eq:deltaep_ipa})  underestimates the true value of  $\delta_{ep}$. This deficiency can be rectified by introducing \textit{enhancement factors} $\gamma_i\geq 1$, which are specific to each electronic molecular orbital, into Eq.~(\ref{eq:deltaep_ipa}), viz.,
\al{\label{eq:conden_general}
 \delta_{ep}
= 2\sum_{i=1}^{N_e/2} \gamma_i \int \left\lvert \varphi_i(\vec{r}) \right\rvert^2  \left\lvert \psi(\vec{r}) \right\rvert^2 \, d\tau .
}
Many-body-theory calculations show (see Refs. \cite{Green15,Green18a}) that the enhancement factors can be approximated by
\al{\label{eq:enhancement_factor}
\gamma_i = 1 + \sqrt{\frac{1.31}{-\epsilon_i}} + \left( \frac{0.834}{-\epsilon_i} \right)^{2.15},
}
where $\epsilon_i<0$ is the energy of electronic orbital $i$.

\subsection{Annihilation from the continuum}\label{subsec:zeff}

Similarly to Eq.~(\ref{eq:Gamma}), the cross section of $2\gamma$ annihilation for a positron incident on a closed-shell target is
\al{
\sigma _a = \pi r_0^2 \frac{c}{v}\Zeff ,
}
where $v$ is the positron velocity and $\Zeff$ is the effective number of electrons that contribute to annihilation. It is given by
\al{\label{eq:zeff}
\Zeff = \int \sum_{i=1}^{N_e} \delta(\vec{r}-\vec{r}_i) \left\lvert\Psi(\vec{r}_1,\dotsc,\vec{r}_{N_e},\vec{r})\right\rvert^2 \, d\tau_1 \dotsm d\tau_{N_e}  d\tau ,
}
which is similar to Eq.~(\ref{eq:deltaep_ipa}) for $\delta _{ep}$, except that the wave function $\Psi(\vec{r}_1,\dotsc,\vec{r}_{N_e},\vec{r})$ now
describes positron scattering by the target. At large positron-target separations,
\al{\label{eq:Zeff_normalization_requirement}
\Psi(\vec{r}_1,\dotsc,\vec{r}_{N_e},\vec{r}) \simeq \Phi(\vec{r}_1,\dotsc,\vec{r}_{N_e})\left[ e^{i\vec{k}\cdot\vec{r}} + f_\text{el}(\vec{k},\vec{k}') \frac{e^{ikr}}{r}\right],
}
where $\Phi(\vec{r}_1,\dotsc,\vec{r}_{N_e})$ is the ground-state wave function of the  target, $\vec{k}$ ($\vec{k}'$) is the momentum of the positron before (after) the collision, and $f_\text{el}(\vec{k},\vec{k}')$ is the elastic scattering amplitude, with $k=k'$ \footnote{Since we treat the target as a static source of potential for the positron, all inelastic scattering channels (except annihilation) are closed for all positron energies.}.

Due to the use of a discrete Gaussian basis, the positive-energy positron pseudostates that we calculate are not \textit{bona fide} scattering states; they decay exponentially, rather than oscillate, at large positron-target separations, and are normalized to unity [see Eq.~(\ref{eq:posnorm})], instead of an asymptotic plane wave, as in Eq.~(\ref{eq:Zeff_normalization_requirement}). However, extraction of the values of $\Zeff$ at the energies of the pseudostates is still possible. As in the calculation of the elastic scattering phase shift, we consider only $s$-type pseudostates.
This means that we calculate only the $s$-wave contribution to $\Zeff$, which dominates at low positron momenta $k$. Here, the contributions of higher partial waves to $\Zeff$ are suppressed as $(kR_a)^{2l}$, where $R_a$ is the radius of the target, so are typically small, unless one of the higher partial waves possesses a shape resonance.
In principle, the method could be extended to compute the contributions to $\Zeff$ from higher partial waves by considering the non-$s$-type pseudostates.

The $s$-wave part of the positron scattering wave function, normalized according to Eq.~(\ref{eq:Zeff_normalization_requirement}), has the asymptotic form
\al{\label{eq:swave_asymp_full}
\psi(\vec{r}) \simeq \frac{\sin(kr+\delta_0)}{kr}.
}
Comparing this with Eqs.~(\ref{eq:spherical_separation}) and (\ref{eq:swave_asymp}),
we see that  $\Zeff$ for the $s$-wave positron can then be found as
\al{\label{eq:Zeff_del}
\Zeff = \frac{4\pi}{A^2} \delta_{ep} ,
}
where $\delta_{ep}$ is the contact density calculated for a positive-energy $s$-type pseudostate, normalized by Eq. (\ref{eq:posnorm}). The normalization constant $A$ 
can be determined by fitting the radial part of the positive-energy pseudostate by the form (\ref{eq:swave_asymp}) in an intermediate range of $r$. In the case of a molecular target, the wave functions of $s$-type pseudostates should also be spherically averaged before fitting (see Appendix \ref{sec:spherical_averaging}).
A similar approach was used in Ref.~\cite{Swann19a} to compute the pickoff annihilation parameter ${}^1\Zeff$ for Ps-atom collisions.

\begin{figure}
\includegraphics[width=3.375in]{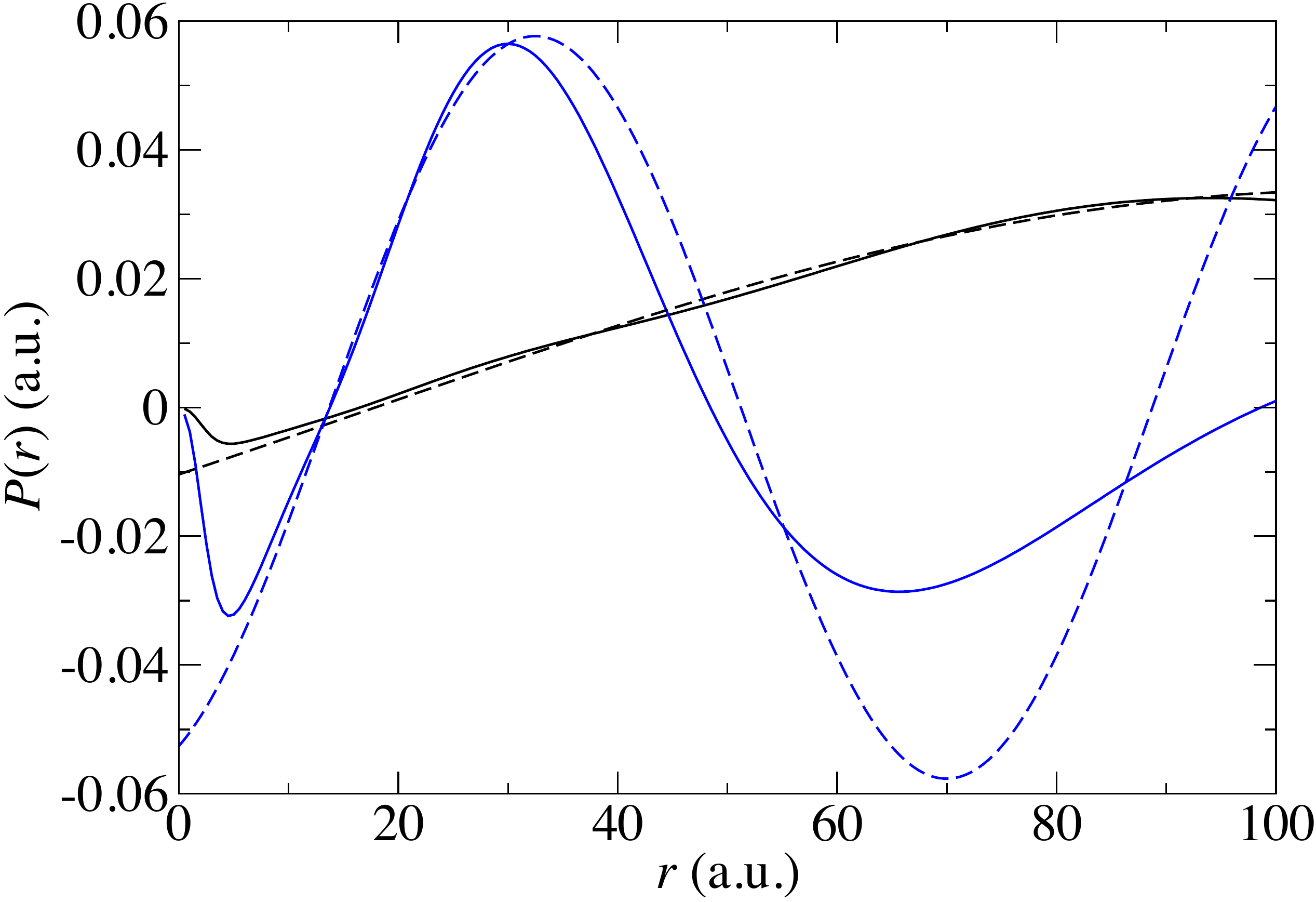}
\caption{\label{fig:Be_radial_12s}Radial functions $P(r)$ for a positron incident on Be, calculated using the $12s$ basis. Solid black line, $P(r)$ for $n=2$ ($\epsilon=1.532\times10^{-4}$~a.u.); solid blue line, $P(r)$ for $n=4$ ($\epsilon=3.507\times10^{-3}$~a.u.); dashed black line, fit for $n=2$ in range $r=10$--50~a.u.; dashed blue line, fit for $n=4$ in range $r=10$--25~a.u.}
\end{figure}

As an example, Fig.~\ref{fig:Be_radial_12s} shows the radial function $P(r)$ for a positron incident on Be, calculated using the $12s$ basis, for the first and third positive-energy pseudostates. (Since Be supports a bound state for the positron, these are actually the $n=2$ and $n=4$ pseudostates, respectively.)
Also shown are fits of  the form (\ref{eq:swave_asymp}). The phase shift $\delta_0$ has been taken as a parameter of the fit. Note that due to the presence of a bound state, the radial functions $P(r)$ have a node at $r\approx 18$~a.u., even for the lowest positive-energy pseudostate ($n=2$). The range of  $r$ used for the fit is chosen in the region between the first and second extrema of $P(r)$. Thus, for $n=2$ we used $r=10$--50~a.u., and for $n=2$, $r=10$--25~a.u. For larger $r$, the decrease of the pseudostate wave function due to the use of the Gaussian basis becomes apparent.

There is, however, an alternative way of determining the values of $A^2$, that does not rely on the fitting of the radial wave function and is free from  related uncertainties. The use of a finite discrete Gaussian basis effectively confines the positron within a soft-walled cavity of (varying) radius $R_0$. In this case, the neighboring positive-energy pseudostates are separated by a momentum difference $\Delta k \approx \pi/R_0$. Away from the target, the wave function of $s$ states takes the form 
\al{
\psi(\vec{r}) = \frac{1}{\sqrt{4\pi}}\frac{A \sin(kr+\delta_0)}{kr}.
}
Assuming that the target size is small compared with $R_0$, the normalization condition (\ref{eq:posnorm}) gives
\al{
 \int_0^{R_0} \frac{A^2 \sin^2(kr+\delta_0)}{(kr)^2} r^2 \, dr = 1.
}
Replacing $\sin ^2(kr+\delta _0)$ with its average value of $\frac12$, we obtain
\al{
A^2 = \frac{2 k^2}{ R_0} \approx \frac{2 k^2\Delta k}{\pi} .
}
Since $k=\sqrt{2\epsilon}$, we have $\Delta k = \Delta \epsilon/\sqrt{2\epsilon}$,
to first order in $\Delta\epsilon$. This can be rewritten as
\al{
\Delta k = \frac{1}{\sqrt{2\epsilon}} \frac{\Delta\epsilon}{\Delta n} \, \Delta n,
}
where $n$ enumerates the pseudostates. Using $\Delta n=1$ for two neighboring pseudostates, and replacing $\Delta\epsilon/\Delta n$ by $d\epsilon/dn$, we have
\al{
\Delta k = \frac{1}{\sqrt{2\epsilon}} \frac{d\epsilon}{dn},
}
so that
\al{\label{eq:A2}
A^2 = \frac{2\sqrt{2\epsilon}}{\pi} \frac{d\epsilon}{dn}.
}
The value of the derivative $d\epsilon / dn$ can be obtained numerically by plotting $\epsilon$ as a function of $n$ and interpolating to real values of $n$ (cf. Sec.~\ref{subsec:elscat} and Fig.~\ref{fig:Be_fn_12s}).

\section{Results}\label{sec:results}

\subsection{\label{sec:IIIA}Atoms: Be, Mg, He, and Ar}\label{subsec:atoms}

We first test the method by computing the binding energy $\eb$ and contact density $\delta_{ep}$ for the positron bound states in Be and Mg. The results are shown in Table \ref{tab:BeMg_binding_contact}, along with a summary of previously published results.
\begin{table}
\caption{\label{tab:BeMg_binding_contact}Binding energy $\eb$ and electron-positron contact density $\delta_{ep}$ for Be and Mg. Brackets indicate powers of 10. Values in bold are the present calculations and those upon which the values of $\rho_A$ were based. The most accurate calculations are denoted (rec.).}
\begin{ruledtabular}
\begin{tabular}{lcc}
 Calculation & $\eb$ (a.u.) & $\delta_{ep}$ (a.u.) \\
\hline
\multicolumn{3}{c}{$e^+$Be calculations} \\
 Present, $12s$ basis & $\boldsymbol{3.090[-3]}$ & $\boldsymbol{9.261[-3]}$ \\
 Present, $19s$ basis & $\boldsymbol{3.129[-3]}$ & $\boldsymbol{9.376[-3]}$ \\
 Stochastic variational \cite{Ryzhikh98} & $1.687[-3]$ & $6.62[-3]$ \\
 Stochastic variational \cite{Mitroy01} & $\boldsymbol{3.147[-3]}$ & $\boldsymbol{8.24[-3]}$ \\
 Configuration interaction \cite{Bromley01} & $3.083[-3]$ & $7.77[-3]$ \\
 Configuration interaction \cite{Bromley06} & $3.169[-3]$ & $8.143[-3]$ \\
 Diffusion Monte Carlo \cite{Mella02} & $1.2[-3]$ & \\
 Stochastic variational (rec.) \cite{Mitroy10} & $3.163[-3]$ & $8.55[-3]$ \\
 Relativistic coupled cluster \cite{Dzuba12} & $6.76[-3]$ \\
  Relativistic coupled cluster \cite{Harabati14} & $7.86[-3]$ \\
  RXCHF\footnotemark[1]~\cite{Brorsen17} & $3.02[-3]$ & $8.12[-3]$ \\
\multicolumn{3}{c}{$e^+$Mg calculations} \\
 Present, $12s$ basis & $\boldsymbol{1.555[-2]}$ & $\boldsymbol{2.365[-2]}$ \\
 Present, $19s$ basis & $\boldsymbol{1.555[-2]}$ & $\boldsymbol{2.368[-2]}$ \\
 Polarized orbital \cite{Szmytkowski93} & $5.5[-4]$ \\
 Polarized orbital \cite{McEachran98} & $4.59[-3]$ \\
 Many-body perturbation theory \cite{Dzuba95} & $3.2[-2]$ \\
 Many-body perturbation theory \cite{Gribakin96} & $3.62[-2]$ \\
 Stochastic variational \cite{Mitroy01} & $\boldsymbol{1.561[-2]}$ & $\boldsymbol{1.89[-2]}$ \\
 Diffusion Monte Carlo \cite{Mella02} & $1.68[-2]$ & \\
 Configuration interaction \cite{Bromley02} & $1.615[-2]$ & $1.8[-2]$ \\
 Configuration interaction (rec.) \cite{Bromley06} & $1.704[-2]$ & $1.962[-2]$ \\
 Relativistic coupled cluster \cite{Dzuba12} & $1.88[-2]$ \\
 Relativistic coupled cluster \cite{Harabati14} & $2.34[-2]$ \\
  RXCHF\footnotemark[1]~\cite{Brorsen17} & $1.19[-2]$ & $1.58[-2]$
\end{tabular}
\end{ruledtabular}
\footnotetext[1]{Reduced explicitly correlated Hartree-Fock}
\end{table}

For Be, changing from the $12s$ basis to the $19s$ basis increases the binding energy by 1.3\% and the contact density by 1.2\%. Such  small changes show that the $12s$ basis is already almost complete. Comparing our $19s$ binding energy with the stochastic-variational calculation of Mitroy and Ryzhikh \cite{Mitroy01}, which was used as the reference in determining the values of $\rho_\text{A}$ \cite{Mitroy02a},  we see that we have agreement at the level of 0.6\%. This very small discrepancy can be ascribed to a slightly different description of the electrostatic field of the Be atom and possibly also due to our basis set not being totally complete.
 The best currently available value of the binding energy is the more recent stochastic-variational calculation \cite{Mitroy10}; both our $12s$ and $19s$ values are in  close agreement with this result. 
 
Regarding the contact density for Be, the present values are 12--14\% larger than those of Mitroy and Ryzhikh \cite{Mitroy01}. The difference is partly because the Hartree-Fock method, which was used to compute the electrostatic potential of the Be atom, underestimates the ionization potential of the atom, i.e., the Hartree-Fock energy of the valence orbital is too small in magnitude, which leads to an overestimate of the enhancement factor from Eq.~(\ref{eq:enhancement_factor}). 
Table \ref{tab:BeMg_contact_contributions} shows the Hartree-Fock energies of the $1s$ and $2s$ orbitals of the Be atom, along with the contribution $\delta_{ep}^{(nl)}$ to the contact density $\delta_{ep}$ from each orbital $nl$, calculated using the $19s$ positron basis without and with the enhancement factors, Eqs.~(\ref{eq:deltaep_ipa}) and (\ref{eq:conden_general}).
\begin{table}
\caption{\label{tab:BeMg_contact_contributions}Contributions $\delta_{ep}^{(nl)}$ to the electron-positron contact density $\delta_{ep}$ from each orbital $nl$ for positron bound states in Be and Mg. Enhancement factors have been calculated using the Hartree-Fock (HF) orbital energies $\epsilon_{nl}$ for the core orbitals, and using either the Hartree-Fock (HF) or experimental (exp.) \cite{CRC} orbital energies $\epsilon_{nl}$ for the valence orbitals.}
\begin{ruledtabular}
\begin{tabular}{ccccc}
Atom & $nl$ & $\epsilon_{nl}$ (a.u.) & \multicolumn{2}{c}{$\delta_{ep}^{(nl)}$ (a.u.)} \\
\cline{4-5}
& & & Unenhanced & Enhanced \\
\hline
Be & $1s$ & $-4.73235$ (HF) & $5.83830[-5]$ & $9.04980[-5]$ \\
     & $2s$ & $-0.309258$ (HF) & $8.07573[-4]$ & $9.28518[-3]$ \\
     &       & $-0.342603$ (exp.) & $8.07573[-4]$ & $7.85547[-3]$ \\
Mg & $1s$ & $-49.0374$ (HF) & $1.65552[-6]$ & $1.92637[-6]$  \\
     & $2s$ & $-3.76634$ (HF) & $9.93184[-5]$ & $1.61777[-4]$ \\
     & $2p$ & $-2.28282$ (HF) & $3.28824[-4]$ & $6.15655[-4]$ \\
     & $3s$ & $-0.253030$ (HF) & $1.40759[-3]$ & $2.28982[-2]$ \\
     &             & $-0.280994$ (exp.) & $1.40759[-3]$ & $1.90445[-2]$
\end{tabular}
\end{ruledtabular}
\end{table}
Also shown is the experimental value of the energy of the valence $2s$ orbital~\cite{CRC}, along with the enhanced value of $\delta_{ep}^{(2s)}$ obtained using this experimental energy. Adding the enhanced contributions to $\delta_{ep}$ of the core $1s$ orbital and the valence $2s$ orbital, where the experimental $2s$ energy has been used to calculate the enhancement factor, gives $\delta_{ep}=7.946\times10^{-3}$~a.u. This is in much better agreement with the value of Mitroy and Ryzhikh \cite{Mitroy01}, at the level of 3.6\%. The remaining discrepancy is partly due to our use Eq.~(\ref{eq:enhancement_factor}) to compute the enhancement factors. If we instead scale the unenhanced contact density  by Mitroy and Ryzhikh's enhancement factors of 2.5 for the $1s$ orbital and 10.18 for the $2s$ orbital \cite{Mitroy01}, we  obtain $\delta_{ep}=8.367\times10^{-3}$~a.u., within 1.5\% of the value in Ref.~\cite{Mitroy01}.

For Mg, the two basis sets give essentially identical results for both the binding energy and the contact density.
The difference in the present binding energy from that of Ref.~\cite{Mitroy01} is 0.4\%. The best currently available calculation by Bromley and Mitroy \cite{Bromley06} gives the binding energy which is 10\% greater than our model-potential value.

Our calculated contact density for Mg is 25\% larger than that of Ref.~\cite{Mitroy01}. Again, this difference is mostly due to an overestimation of the enhancement factor for the valence $3s$ orbital when using the Hartree-Fock orbital energy. Table \ref{tab:BeMg_contact_contributions} shows the Hartree-Fock energy of each orbital of the Mg atom, along with the contribution $\delta_{ep}^{(nl)}$ to the contact density $\delta_{ep}$ from each orbital, as calculated using the $19s$ positron basis, without and with the enhancement factors.
Also shown is the experimental value of the energy of the valence $3s$ orbital~\cite{CRC}, along with the enhanced contribution of the $3s$ orbital to $\delta_{ep}$, obtained using this energy.
 Adding the enhanced contribution to $\delta_{ep}$ from the core  orbitals to that from the valence $3s$ orbital, where the experimental $3s$ energy has been used to calculate the enhancement factor, gives $\delta_{ep}=1.982\times10^{-2}$~a.u., which is within 4.9\% of the value of Mitroy and Ryzhikh \cite{Mitroy01}. 
Again, the remaining discrepancy may be partly due to our use of Eq.~(\ref{eq:enhancement_factor}) for the enhancement factors. If we instead scale the unenhanced contact density by Mitroy and Ryzhikh's enhancement factors of 2.5 for the core orbitals and 13.2 for the $3s$ orbital \cite{Mitroy01}, we  obtain $\delta_{ep}=1.965\times10^{-2}$~a.u., which reduces the discrepancy to 4.0\%. Interestingly, the latter value of $\delta _{ep}$ is in near-exact agreement with the recommended configuration-interaction value of $1.962\times10^{-2}$~a.u.~\cite{Bromley06}.

\begin{figure*}
\centering
\includegraphics[width=.9\textwidth]{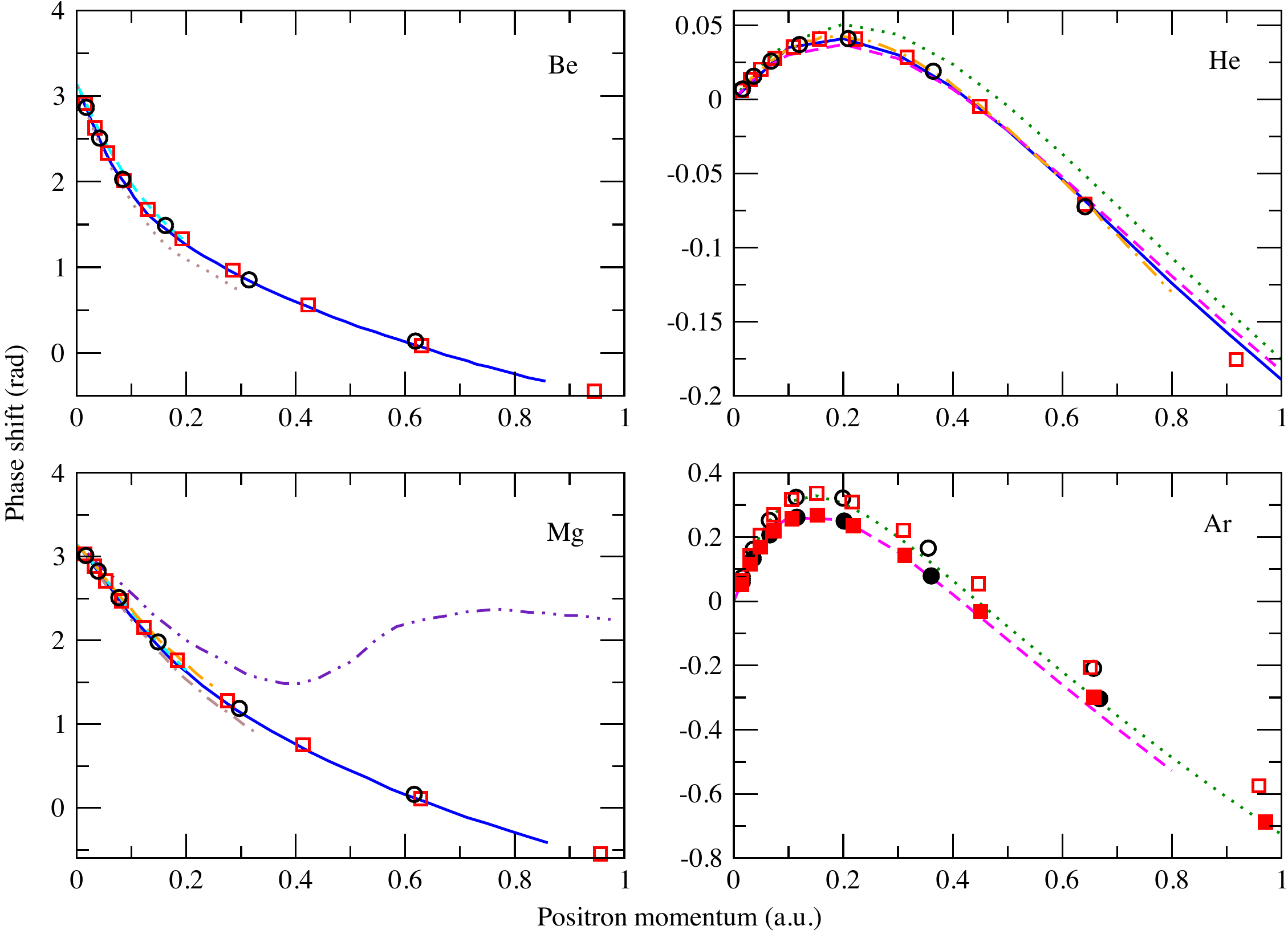}
\caption{\label{fig:BeMgHeAr_phase_shifts}Calculations of the $s$-wave phase shift for elastic scattering of a positron by Be, Mg, He, and Ar. Black circles, present calculations using the $12s$ positron basis; red squares, present calculations using the $19s$ positron basis; for Ar, open (filled) symbols are for $\rho_\text{Ar}=1.710$~a.u. ($\rho_\text{Ar}=1.88$~a.u.) Lines are results of existing calculations, as follows. For Be and Mg: solid blue lines, model potential \cite{Bromley98}; short-dashed cyan lines, model potential \cite{Poveda16}; 
dotted brown line (Be only), effective single-particle potential \cite{Zubiaga14}; long-dashed orange line (Mg only), model potential \cite{Mitroy08}; dot-dashed brown line (Mg only), close coupling with model potential \cite{Savage11}; dot-dash-dotted indigo line (Mg only), $\operatorname{Re}\delta_0$ from many-body theory \cite{Gribakin96}. For He and Ar: dotted green lines, polarized orbital \cite{McEachran78,McEachran79}; short-dashed magenta lines, many-body theory \cite{Green14}; solid blue line (He only), Kohn variational \cite{vanReeth99}; dot-dashed orange line (He only), model potential \cite{Mitroy02a}.}
\end{figure*}

We next consider positron elastic scattering from Be, Mg, He, and Ar.
Figure \ref{fig:BeMgHeAr_phase_shifts} shows the $s$-wave phase shifts for the four atoms, obtained from Eq.~(\ref{eq:delta0}), as functions of the positron momentum. 
For each of the atoms, the phase shifts obtained using the $12s$ and $19s$ positron basis sets are very close.
Figure~\ref{fig:BeMgHeAr_phase_shifts} also shows results of several existing calculations.

For Be and Mg, most of the existing calculations use model potentials \footnote{There are also accurate convergent close-coupling calculations for Mg \cite{Utamuratov12}, which unfortunately, do not report the phase shifts.}. We obtain near-exact agreement with the calculation of Bromley \textit{et al.} \cite{Bromley98}, which used the same model as the present calculations, with just slightly different values of the cutoff radii: $\rho_\text{Be}=2.7084$~a.u. and $\rho_\text{Mg}=3.0720$~a.u. We obtain excellent agreement with the other calculations, except the many-body-theory calculation for Mg of Gribakin and King \cite{Gribakin96}; however, that calculation overestimated the attractive virtual-Ps-formation component of the many-body correlation potential, which resulted in a larger binding energy (see Table \ref{tab:BeMg_binding_contact}) and higher phase shifts. Also, this is the only calculation shown that incorporated the Ps-formation channel which is open for $k>0.25$~a.u., making the phase shift complex and leading to a rapid momentum dependence of $\operatorname{Re}\delta_0$.

For He, we have very close agreement with the calculation of Mitroy and Ivanov \cite{Mitroy02a} that used the same model potential.
We also obtain excellent agreement with the near-exact Kohn-variational calculation of van Reeth and Humberston \cite{vanReeth99} (which was used in Ref.~\cite{Mitroy02a} to choose the value of $\rho_\text{He}$) and the recent many-body-theory calculation of Green \textit{et al.} \cite{Green14}. For Ar, the calculations with $\rho_\text{Ar}=1.710$~a.u. closely follow the polarized-orbital calculation of McEachran \textit{et al.} \cite{McEachran79} (which was the reference calculation for choosing
$\rho_\text{Ar}$ in Ref.~\cite{Mitroy02a}). On the other hand, using $\rho_\text{Ar}=1.88$~a.u., we reproduce the more advanced many-body-theory calculation of Green \textit{et al.} \cite{Green14}  across the energy range considered.

Table \ref{tab:atom_scattering_lengths} shows the values of the scattering length obtained for Be, Mg, He, and Ar, using the $12s$ and $19s$ basis sets, along with a selection of existing calculations and an experimental datum for Ar.
\begin{table}
\caption{\label{tab:atom_scattering_lengths}Positron scattering lengths in a.u. for Be, Mg, He, and Ar, from the present $12s$ and $19s$ and other calculations, and experiment.}
\begin{ruledtabular}
\begin{tabular}{ccccc}
 & Be & Mg & He & Ar \\
 \hline
$12s$ & 14.98 & 7.397 & $-0.5017$ & $-5.786$\footnotemark[1], $-4.758$\footnotemark[2]  \\
$19s$ & 13.61 & 6.709 & $-0.5056$ & $-5.797$\footnotemark[1], $-4.764$\footnotemark[2] \\
Other  & 15.6~\cite{Mitroy02a} & 6.76~\cite{Mitroy02a} & $-0.481$~\cite{Mitroy02a} & $-5.29$\footnotemark[1]~\cite{Mitroy02a} \\
          & 13.3~\cite{Poveda16} & 6.2~\cite{Poveda16} & $-0.529$~\cite{McEachran78} & $-5.30$~\cite{McEachran79}  \\
         & 13.8~\cite{Reid14} & 4.2~\cite{Gribakin96} & $-0.48$~\cite{Campeanu77} & $-4.3$~\cite{Fursa12} \\
         & 18.76~\cite{Zubiaga14} & 7.23~\cite{Mitroy08} & $-0.435$~\cite{Green14} & $-4.41$~\cite{Green14} \\
         & 16~\cite{Bromley98} & 8.5~\cite{Utamuratov12} & $-0.474$~\cite{Zhang08} & \\
         & & & $-0.452$~\cite{Zhang11} & \\
         & & & $-0.45$~\cite{Poveda13} & \\
Exp. & & & & $-4.9\pm0.7$~\cite{Chiari14}      
\end{tabular}
\end{ruledtabular}
\footnotetext[1]{$\rho_\text{Ar}=1.710$~a.u.}
\footnotetext[2]{$\rho_\text{Ar}=1.88$~a.u.}
\end{table}
Note that for He and Ar, which do not bind the positron, the scattering length has been calculated via Eq.~(\ref{eq:Be_scatlength_fit_b}) using the second and third positive-energy pseudostates rather than the first and second positive-energy pseudostates.
 This is because for the first positive-energy pseudostate ($n=1$), we have $\Delta\epsilon_n<0$, which means that to calculate the phase shift for this pseudostate, one has to extrapolate $g(\ln\epsilon)$ to a value of $\epsilon$ that is smaller than the lowest free-particle energy, $\epsilon_1^{(0)}$. Such extrapolation is less reliable than interpolation to values of $\epsilon$ that are within the range of  the free-particle energies.
 
For both Be and Mg, the $12s$ value of the scattering length is 10\% larger than the $19s$ value. For He and Ar, the difference between the $12s$ and $19s$ values is much smaller, less than 1\%.
Broadly speaking, there is reasonably good agreement with the existing calculations for all four atoms. We can make a direct comparison with the results of Ref.~\cite{Mitroy02a}, where the same model was used as in the present calculations (taking $\rho_\text{Ar}=1.710$~a.u.).
Our $19s$ values of the scattering length for Be, Mg, He, and Ar are in agreement with those of Ref.~\cite{Mitroy02a} at the level of 13\%, 0.75\%, 5.1\%, and 9.6\%, respectively. The discrepancies are due to the different method we have used to extract the scattering length: we have used an effective-range-theory fit to the $s$-wave phase shift, while in Ref.~\cite{Mitroy02a} it is inferred from the zero-energy elastic scattering cross section
\footnote{The scattering length for the calculations of Mitroy and Ivanov \cite{Mitroy02a} can be determined from the zero-energy elastic cross sections $\sigma$  given in Table II of Ref.~\cite{Mitroy02a} (in units of $\pi a_0^2$, where $a_0$ is the Bohr radius) by $a=\pm \sqrt{\sigma}/2$.}.
As expected, the scattering length for Be and Mg is large and positive (since they support weakly bound states for the positron), while for He and Ar it is negative. For Ar it is quite large in magnitude. Using our $19s$ value of the scattering length for $\rho_\text{Ar}=1.88$~a.u., $a=-4.76$~a.u., we estimate the energy of the virtual level to be $\epsilon\approx 1/2a^2=0.022$~a.u. This scattering length is in perfect agreement with experiment \cite{Chiari14}, though the latter has relatively large error bars.

\begin{figure*}
\centering
\includegraphics[width=0.9\textwidth]{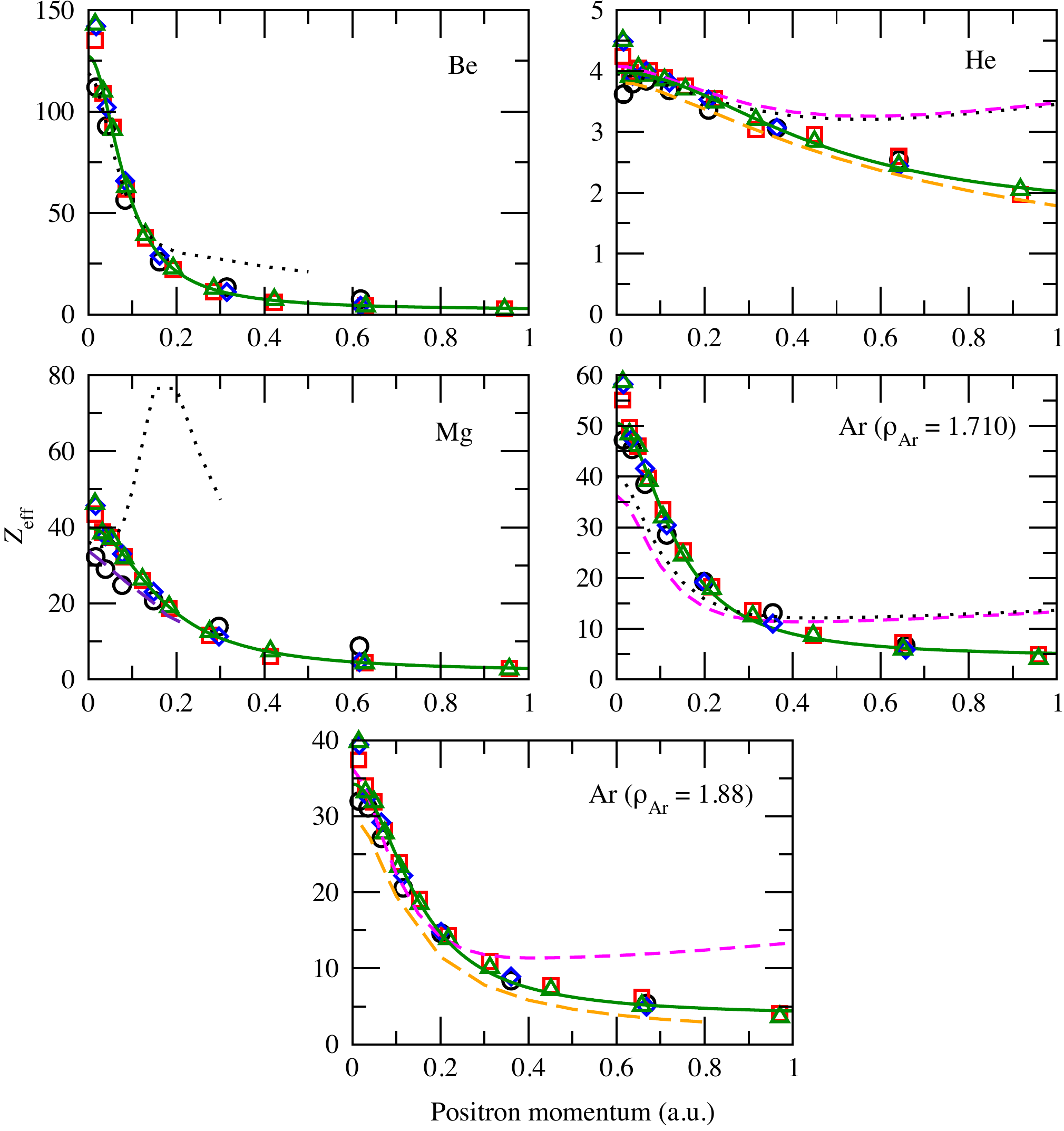}
\caption{\label{fig:BeMgHeAr_Zeff}Calculations of  $\Zeff$ for Be, Mg, He, and Ar. Black circles, $12s$ calculation of $s$-wave $\Zeff$ using radial fit (\ref{eq:swave_asymp}) for normalization; red squares, $19s$ calculation of $s$-wave $\Zeff$ using radial fit for normalization; blue diamonds, $12s$ calculation of $s$-wave $\Zeff$ using analytical estimate (\ref{eq:A2}) for normalization; green triangles, $19s$ calculation of $s$-wave $\Zeff$ using analytical estimate for normalization. Solid green lines are fits of the form (\ref{eq:Zeff_k_dep}) to the $19s$ calculation using analytical estimate for normalization.
Other lines are results of existing calculations, as follows:  short-dashed magenta lines, polarized orbital (total $\Zeff$) \cite{McEachran78,McEachran79}; long-dashed orange lines, many-body theory ($s$-wave $\Zeff$) \cite{Green14}; long-dashed indigo line, model potential ($s$-wave $\Zeff$) \cite{Mitroy08}, dotted black lines, model potential (total $\Zeff$) \cite{Mitroy02a}.}
\end{figure*}

Finally, we compute the annihilation parameter $\Zeff$ for Be, Mg, He, and Ar.
Figure \ref{fig:BeMgHeAr_Zeff} shows the results, along with several previous calculations, some of which reported  the $s$-wave component of $\Zeff$ (which is what we calculate), while others reported the total $\Zeff$. For each target atom, we show values of $\Zeff$ obtained from Eq.~(\ref{eq:Zeff_del}) with either the $12s$ or $19s$ positron basis, and using the normalization factor $A^2$ from either the radial fit (\ref{eq:swave_asymp}) (as in Fig.~\ref{fig:Be_radial_12s}) or from Eq.~(\ref{eq:A2}). For Ar, the results for $\rho_\text{Ar}=1.710$~a.u. and $\rho_\text{Ar}=1.88$~a.u. are presented in separate panels.

Considering $\Zeff$ as a function of $k$, our results are largely independent of the positron basis set used, and of the method of normalization of the positron wave function. The most significant discrepancies occur for small $k$. In fact, with the exception of the $12s$ basis set, the value of $\Zeff$ that comes from the first positive-energy pseudostate for each atom appears to be an outlier that sits above  the trend set by the other pseudostates with small momenta. The exact reason for this behavior is unclear, but it may relate to the uncertainty in normalizing the lowest-energy pseudostate correctly.
The $12s$ basis with normalization determined by radial fit appears to yield smaller values of $\Zeff$ near $k=0$ when compared to the other calculations.

Figure \ref{fig:BeMgHeAr_Zeff} shows that for all atoms except He, the $\Zeff$ is strongly enhanced at low positron momenta. This occurs when the positron-atom potential supports a weakly bound or low-lying virtual $s$ level. In this case the momentum dependence of $\Zeff$ at low momenta $k$ has the form \cite{Goldanskii64,Dzuba93,Dzuba96,Gribakin00,Green14}
\al{\label{eq:Zeff_k_dep}
\Zeff =  \frac{F}{\kappa^2+k^2} +B,
}
where $B$ and $F$ are constants. The first term, in which $\kappa\approx1/a$ ($\lvert\kappa\rvert \ll1$),
with $a$ the scattering length, is due to the $s$-wave component of the positron wave function being ``in resonance'' with the bound or virtual level. The constant $B$ accounts for the nonresonant background $\Zeff$, which depends weakly on $k$. 
Figure \ref{fig:BeMgHeAr_Zeff} shows fits of the form (\ref{eq:Zeff_k_dep}) for the calculations that used the $19s$ positron basis set and the analytical estimate for the normalization of the positron wave function; the values of the fitting parameters are given in Table \ref{tab:Zeff_atoms}. Note that the outlying value of $\Zeff$ for the first positive-energy pseudostate was ignored when determining the fitting parameters.
\begin{table}
\caption{\label{tab:Zeff_atoms}Fitting parameters for $\Zeff$ for Be, Mg, He, and Ar, as calculated using the $19s$ positron basis set and the analytical estimate for the normalization of the positron wave function.}
\begin{ruledtabular}
\begin{tabular}{lccc}
Atom & $\kappa$ & $F$ & $B$ \\
\hline 
Be & 0.0853 & $0.910$  & 1.99 \\
Mg & $0.167$ & $1.06$ & 1.88 \\
He & $-0.456$ & 0.493 & 1.61 \\
Ar ($\rho_\text{Ar}=1.710$) & $-0.134$ & 0.830 & 4.32 \\
Ar ($\rho_\text{Ar}=1.88$) & $-0.149$ & 0.682 & 3.73 
\end{tabular}
\end{ruledtabular}
\end{table}

As expected, the atom with the largest threshold value of $\Zeff$ is Be ($\Zeff\approx 127$ for the 19s calculation), which has the largest absolute value of the scattering length, i.e., the smallest $\kappa$. The He atom has the smallest threshold value of $\Zeff$ and, overall, the weakest dependence of $\Zeff$ on $k$.

For all atoms, there is broadly good agreement with the model-potential calculations of Mitroy and Ivanov \cite{Mitroy02a} for low momenta. (For Ar, we only compare the results of Ref.~\cite{Mitroy02a} with the present calculations for $\rho_\text{Ar}=1.710$~a.u.) Note that the results of Ref.~\cite{Mitroy02a} are for the total $\Zeff$, not just the $s$-wave component: at higher momenta, the results of Ref.~\cite{Mitroy02a} become significantly larger than the present results due to the contribution of  higher partial waves. This is particularly conspicuous for Mg, where the total $\Zeff$ is strongly peaked at $k\approx0.2$~a.u. due to a $p$-wave shape resonance \cite{Mitroy02a}. We note that at $k=0$, all of the present values of $\Zeff$, except for the $12s$ calculation with the radial-fit normalization, are 10--20\% larger than those predicted by Ref.~\cite{Mitroy02a}. This is at least partly due to the enhancement factors used in the present work being larger than those used by Mitroy and Ivanov \cite{Mitroy02a}.

For He, we observe excellent agreement at low momenta with the polarized-orbital calculation of total $\Zeff$ of McEachran \textit{et al.}~\cite{McEachran78} and with the many-body-theory calculation of $s$-wave $\Zeff$ \cite{Green14}. 
For Ar, the agreement with Ref.~\cite{McEachran78} is better for $\rho_\text{Ar}=1.88$~a.u. than for $\rho_\text{Ar}=1.710$~a.u. For $\rho_\text{Ar}=1.88$~a.u. we also have good agreement with the many-body-theory calculation of $s$-wave $\Zeff$ \cite{Green14}, though the present results are slightly larger. For Mg, our $12s$ calculation with normalization determined using the radial fit is in excellent agreement with the model-potential calculation of $s$-wave $\Zeff$ of Mitroy \textit{et al.}~\cite{Mitroy08}, but once more, our other calculations give somewhat larger values.

Overall, our calculations show that positive-energy square-integrable pseudostates can be used to determine the $s$-wave scattering phase shifts and the normalized  rates $\Zeff$ for positron in-flight annihilation at low energies.

\subsection{Molecules: H$_2$, N$_2$,  Cl$_2$, and CH$_4$}

We now turn our attention to positron interactions with small molecules. We first consider H$_2$, for which there has already been a significant amount of theoretical and experimental investigation of scattering and annihilation. We will examine the dependence of the $s$-wave scattering phase shift and $\Zeff$ on the choice of the positron basis, and the sensitivity of $\Zeff$ to the method used to normalize the positron pseudostates.

\begin{figure}
\centering
\includegraphics[width=3.375in]{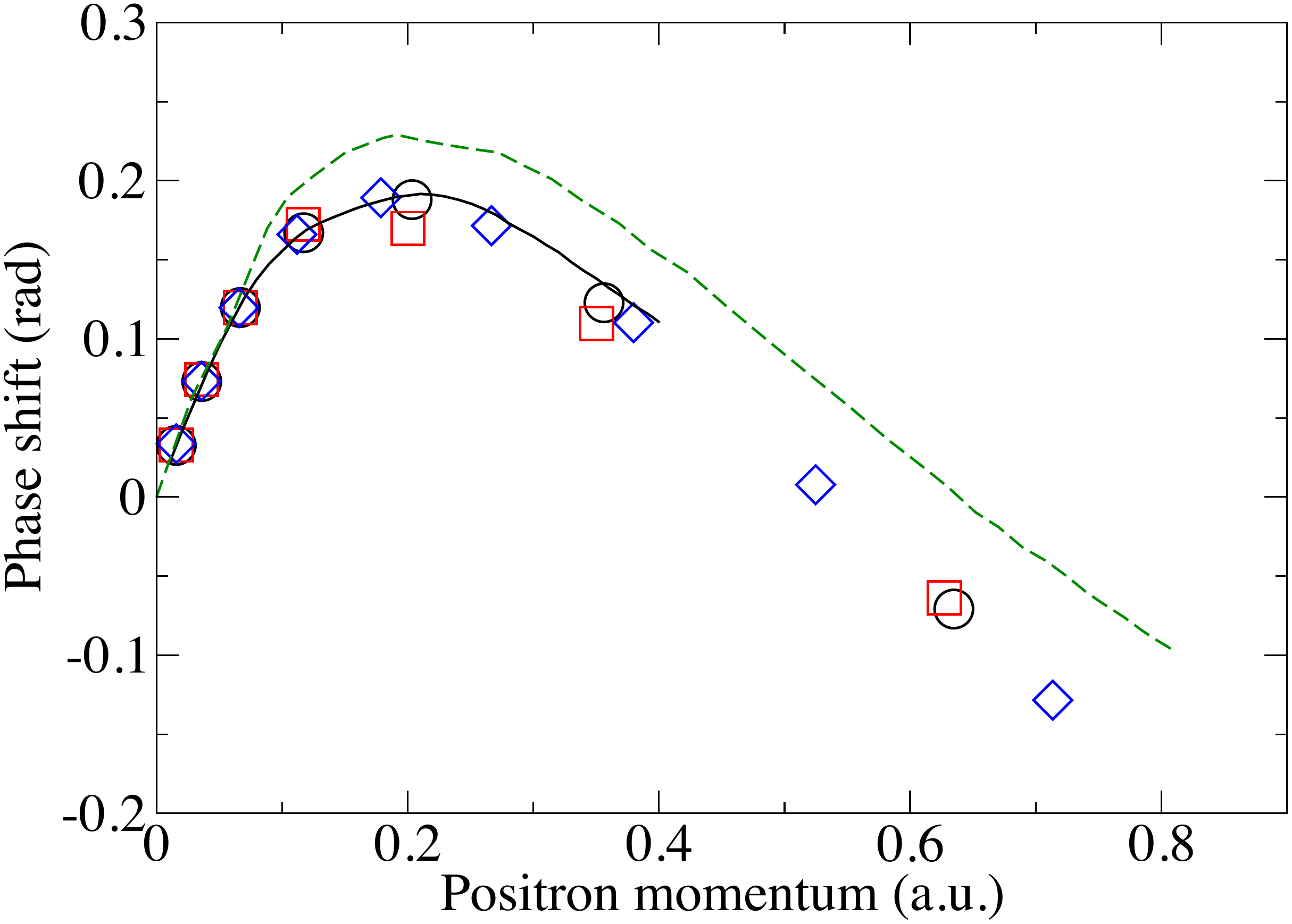}
\caption{\label{fig:H2_phase}Calculations of the $s$-wave phase shift for elastic scattering of a positron by H$_2$. Black circles, $12s$ positron basis; red squares, $12s\,8p$ positron basis; blue diamonds, $12s\,8p\,8d$ positron basis. Solid black line, Kohn-variational calculation of Cooper \textit{et al.} \cite{Cooper08}; green dashed line, modified-effective-range-theory fit of the measured cross section by Fedus \textit{et al.} \cite{Fedus15}.}
\end{figure}

Figure \ref{fig:H2_phase} shows the $s$-wave phase shift for H$_2$, obtained from Eq.~(\ref{eq:delta0})  using the $12s$, $12s\,8p$, and $12s\,8p\,8d$ positron basis sets. These data correspond to the internuclear distance of $R=1.39$~a.u.
Overall, there is little difference between the three sets of results. However, at $k\approx0.2$~a.u., the $12s\,8p$ calculation gives a smaller phase shift than the others. We note that this occurs for the fifth data point, which is where the $p$-type Gaussians start to contribute. The figure also shows the Kohn-variational calculation of Cooper \textit{et al.} \cite{Cooper08}, using the trial scattering wave function that is referred to as $\Psi_t^{(1,A)}$ in Ref.~\cite{Cooper08}.
The present calculations are in near-perfect agreement with those of Cooper \textit{et al.} \cite{Cooper08} at small momenta. This is clear evidence of the ability of the present method to accurately describe low-energy scattering of positrons by small molecules. Finally, the figure shows the phase shift of Fedus \textit{et al.} \cite{Fedus15}, which was determined empirically by performing a fit based on modified effective-range theory to experimental data of Machacek \textit{et al.} \cite{Machacek13} on the $e^+$-H$_2$ elastic  scattering cross section. The result of this fit is very close to the present calculation for positron momenta $k\lesssim0.1$~a.u., but lies slightly higher at larger $k$.

For the $e^+$-H$_2$ scattering length, the $12s$, $12s\,8p$, and $12s\,8p\,8d$ calculations give values of $a=-2.38$, $-2.48$, and $-2.39$~a.u., respectively.
As was the case for the atomic targets that did not support a bound state for the positron, we have calculated these values using the second and third positive-energy pseudostates.
While the $12s$ and $12s\,8p\,8d$ basis sets give almost exactly the same value for the scattering length, the $12s\,8p$ gives a value 4\% larger in magnitude.
 The $12s\,8p\,8d$ value of $a=-2.39$~a.u. should be considered the most reliable. It is close to the stochastic-variational calculations, which gave $a=-2.6$ ($R=1.40$~a.u.)  \cite{Zhang09}, $a=-2.71$ ($R=1.45$~a.u.)  \cite{Zhang11}, and $a=-2.79$~a.u. ($R=1.45$~a.u.)  \cite{Zhang14}, and
is in good agreement with the convergent-close-coupling calculation of Zammit \textit{et al.}, which gave  $a=-2.49$~a.u. ($R=1.40$~a.u.) \cite{Zammit13}.
We also note the existence of an $R$-matrix calculation of Zhang \textit{et al.}, which gave $a=-2.06$~a.u. ($R=1.40$~a.u.) \cite{Zhang11a}.

%
%
%
%

\begin{figure}[ht!]
\centering
\includegraphics[width=3.375in]{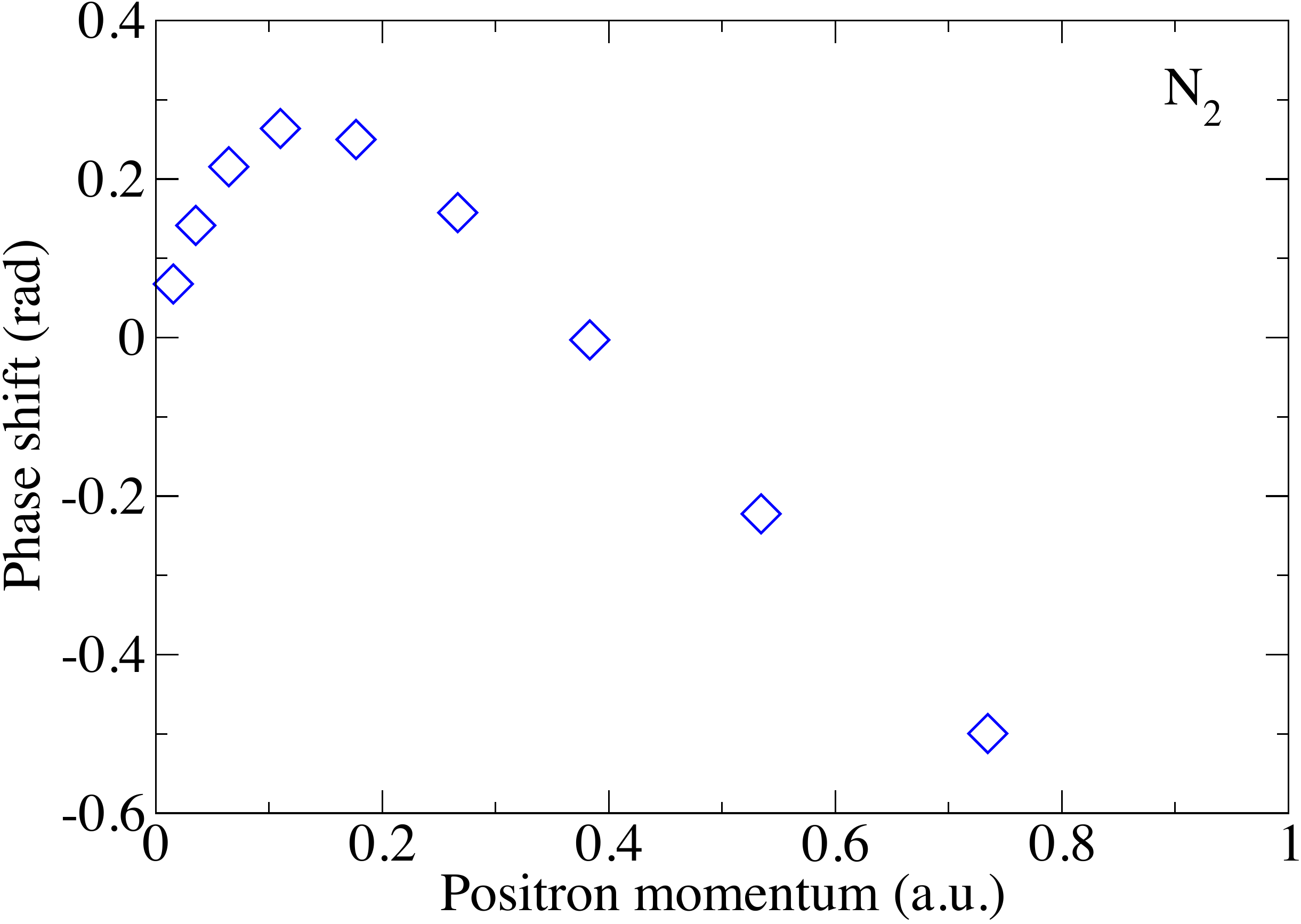} \\[0.5em]
\includegraphics[width=3.375in]{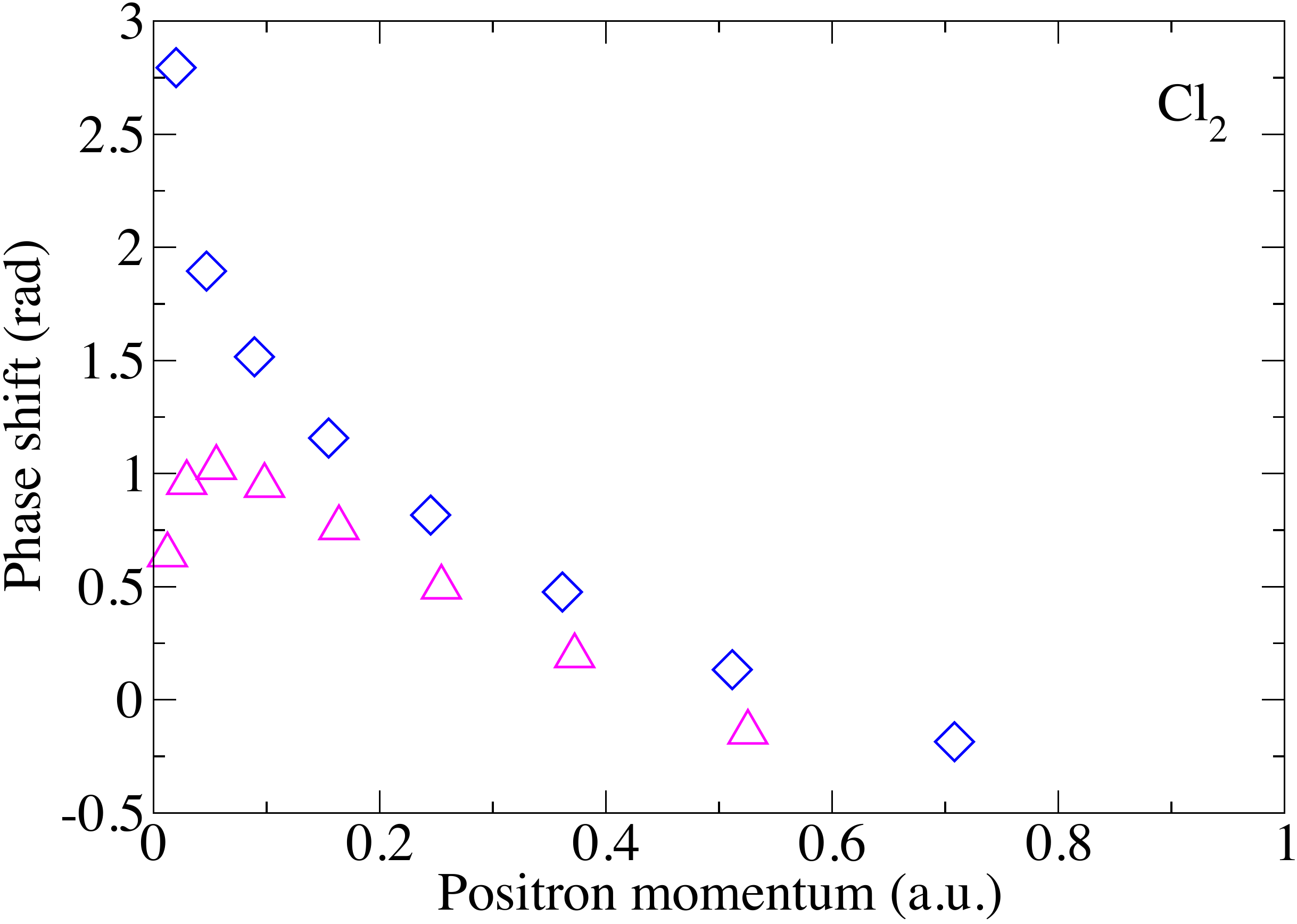} \\[0.5em]
\includegraphics[width=3.375in]{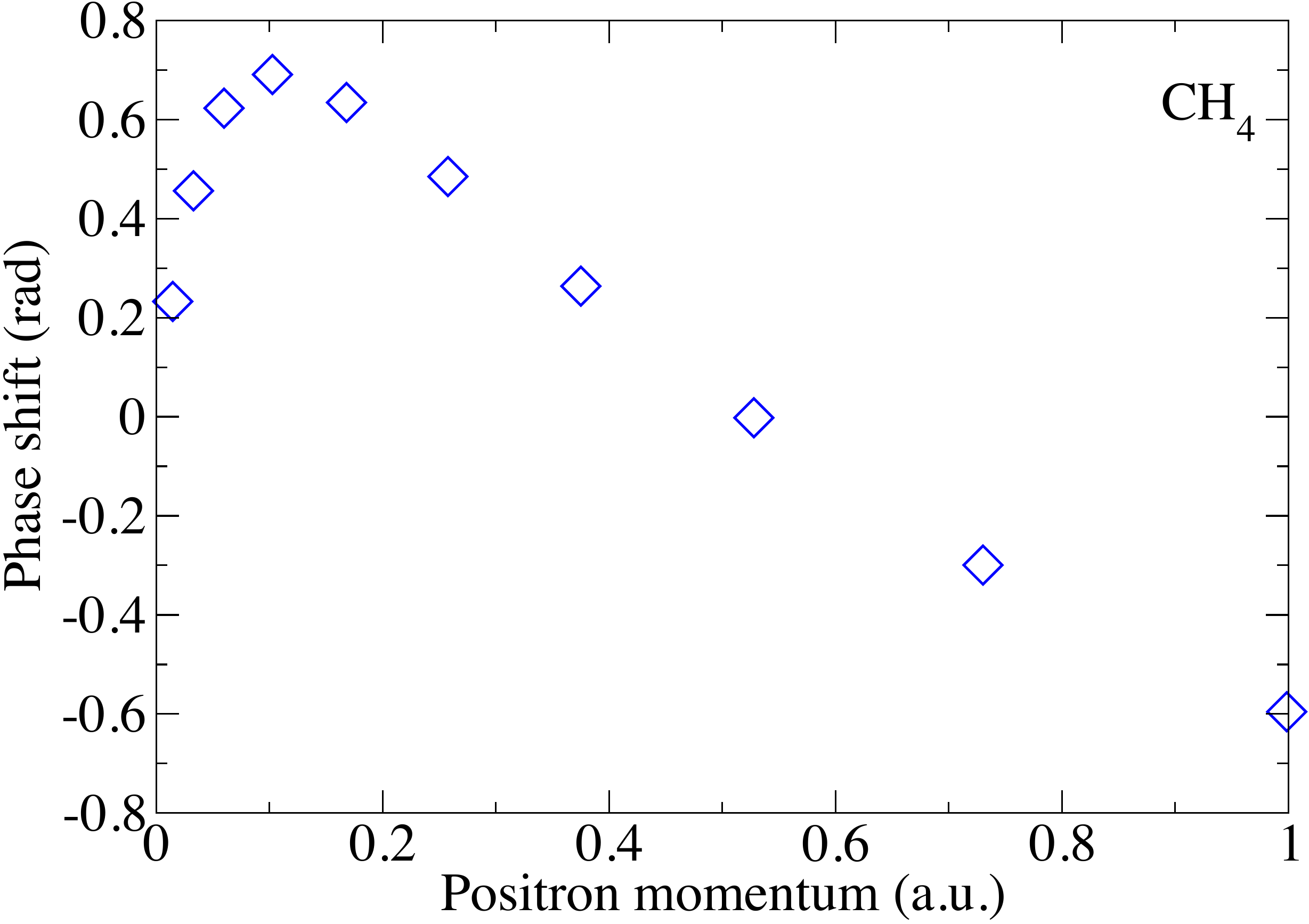}
\caption{\label{fig:N2Cl2CH4_phase}Positron $s$-wave scattering phase shift for N$_2$, Cl$_2$, and CH$_4$, calculated using the  $12s\,8p\,8d$ positron basis for N$_2$ and Cl$_2$ and the $12s\,8p\,8d\,/\,8s$ positron basis for CH$_4$. For Cl$_2$, blue diamonds are for $\rho_\text{Cl}=1.88$~a.u., while magenta triangles are for $\rho_\text{Cl}=2.20$~a.u.}
\end{figure}

Figure \ref{fig:N2Cl2CH4_phase} shows the $s$-wave phase shift for N$_2$, Cl$_2$, and CH$_4$, obtained using the $12s\,8p\,8d$ basis for N$_2$ and Cl$_2$ and the $12s\,8p\,8d\,/\,8s$ basis for CH$_4$. These molecules are more polarizable and more attractive for the positron than H$_2$. The scattering calculations for these targets indicate the presence of virtual states (N$_2$, CH$_4$, and Cl$_2$ with $\rho_\text{Cl}=2.20$~a.u.), or possibly even a weakly bound state (Cl$_2$, $\rho_\text{Cl}=1.88$~a.u.).

For N$_2$, we obtain the scattering length by extrapolating $\tan \delta _0/k$ towards $k=0$, and find $a=-4.6\pm 0.1$~a.u. (The error bars reflect the uncertainty of effective-range-type extrapolation procedure.) Positron scattering from N$_2$ is similar to that from Ar. This could be expected, since the dipole polarizability of N$_2$ is $\alpha = 12.9$~a.u., which is close to 11.1~a.u. for Ar.

For Cl$_2$, our calculations with the smaller cutoff radius $\rho_\text{Cl}=1.88$~a.u.,
predict  a weakly bound state with  $\epsilon_b=4.004\times10^{-4}$~a.u. and  $\delta_{ep}=5.391\times10^{-3}$~a.u. We can then estimate the scattering length as $a\approx1/\sqrt{2\epsilon_b}= 35.5$~a.u. Examining the behaviour of $k\cot \delta _0$ suggests a noticeable uncertainty in the value of the lowest-energy phase shift $\delta _0$ (that is hard to discern on the scale of Fig.~\ref{fig:N2Cl2CH4_phase}). We thus use the second and third pseudostates for extrapolation and find the scattering length $a=26$~a.u. The discrepancy with the above value is related to the fact that for large scattering lengths, the validity of the effective-range expansion, Eq.~(\ref{eq:mert}), is restricted to a narrow range of momenta, $k\lesssim 1/\lvert a \rvert$. This suggests that the scattering length estimated from the binding energy is more reliable.


For Cl$_2$ with $\rho_\text{Cl}=2.20$~a.u., the positron-molecule potential is not attractive enough to support a bound state. However, the very rapid growth of the $s$-wave phase shift at small momenta indicates the presence of a low-lying virtual $s$ level. Using linear extrapolation of $\tan \delta _0/k$ towards $k=0$ gives $a=-65\pm 5$~a.u, where the errors bars again reflect the uncertainty of the extrapolation procedure.
This large negative scattering lengths indicates a virtual level with the energy $\epsilon\approx1/2a^2\sim 10^{-4}$~a.u.

For CH$_4$, we estimate that the scattering length to be $a=-17\pm 1$~a.u. This indicates the presence of a low-lying virtual level, with energy $\epsilon\approx1/2a^2=1.5\times10^{-3}$~a.u. A recent Schwinger multichannel calculation of the $e^+$-CH$_4$ scattering length by Zecca \textit{et al.} gave $a=-7.4$~a.u. \cite{Zecca12}. This is noticeably smaller in magnitude than the present estimate, indicating weaker positron-target attraction (possibly as a result of an incomplete account of the difficult virtual Ps contribution). An older semiempirical calculation by Frongillo \textit{et al.} gave $a=-13.0$~a.u. \cite{Frongillo94}, which is closer to our estimate.

%

\begin{figure}
\centering
\includegraphics[width=3.375in]{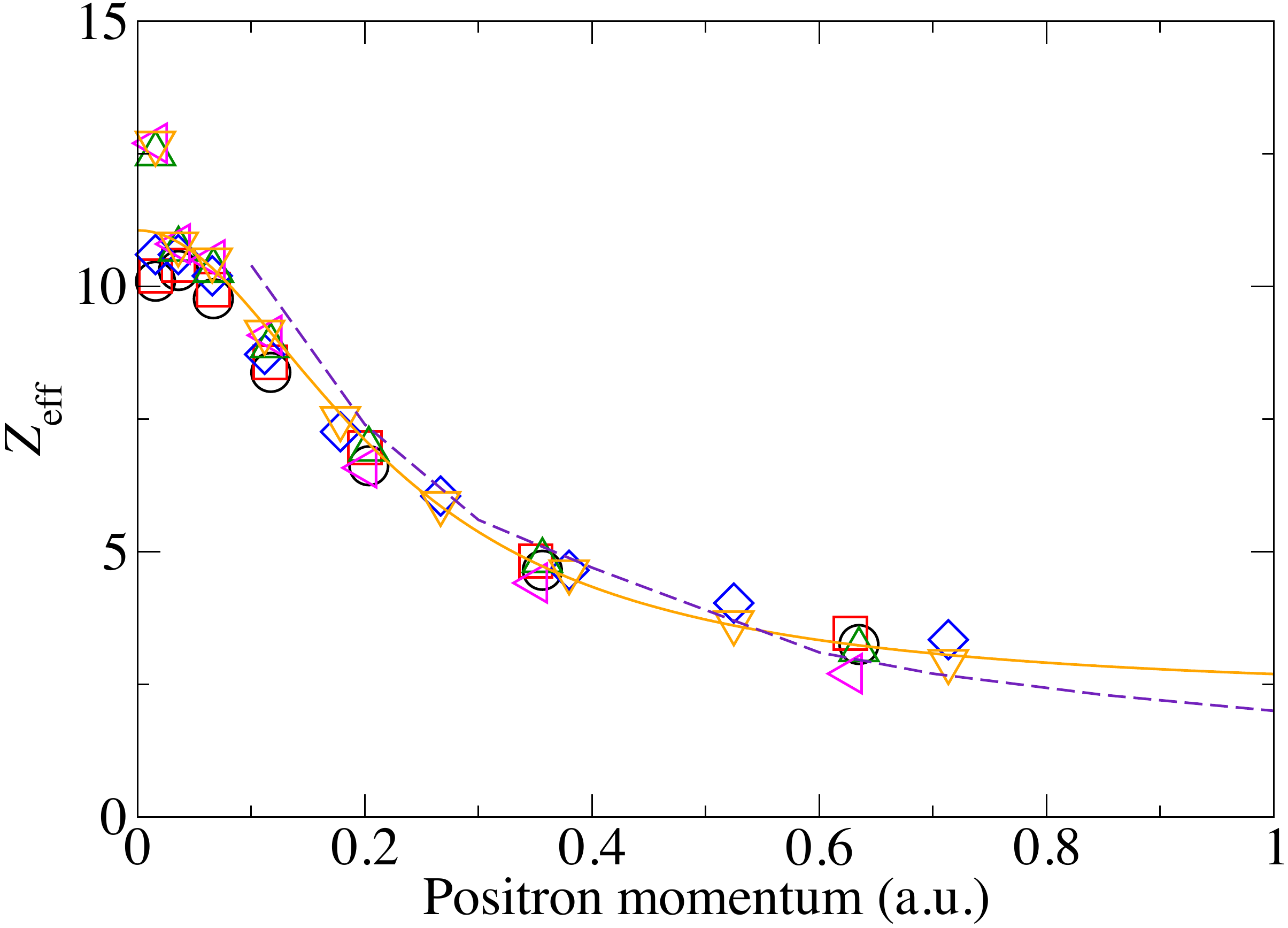}
\caption{\label{fig:H2_Zeff}Calculated $s$-wave $\Zeff$ for H$_2$. Black circles, $12s$ basis; red squares, $12s\,8p$ basis; blue diamonds, $12s\,8p\,8d$ basis (all using radial fit for normalization). Green up triangles, $12s$ basis; magenta left triangles, $12s\,8p$ basis; orange down triangles; $12s\,8p\,8d$ basis (all using analytical estimate, Eq.~(\ref{eq:A2}), for normalization). The solid orange line is a fit of the form (\ref{eq:Zeff_k_dep}) to the $12s\,8p\,8d$ calculation using analytical estimate for normalization. The dashed purple line is the Kohn-variational calculation of Armour and Baker \cite{Armour86}.} 
\end{figure}

Figure \ref{fig:H2_Zeff} shows $\Zeff$ as a function of the positron momentum for H$_2$.
As was the case for the atomic targets, the results are largely independent of the choice of positron basis set and the method of determining the normalization of the positron wave function. However, for all three basis sets, the lowest  positive-energy pseudostate appears to give values of $\Zeff$ that are too large when the analytical estimate for the normalization, Eq.~(\ref{eq:A2}), is used.
The figure also shows a fit of the form (\ref{eq:Zeff_k_dep}) to the  $12s\,8p\,8d$ calculation using the analytical estimate for the normalization, with the first pseudostate excluded from the fit; the fitting parameters are given in Table \ref{tab:Zeff_molecules}.
\begin{table}
\caption{\label{tab:Zeff_molecules}Fitting parameters for $\Zeff$ for H$_2$, N$_2$, Cl$_2$, and CH$_4$, as calculated using the $12s\,8p\,8d$ positron basis set for H$_2$, N$_2$, and Cl$_2$, and the $12s\,8p\,8d\,/\,8s$ positron basis set for CH$_4$, and the analytical estimate for the normalization of the positron wave function.}
\begin{ruledtabular}
\begin{tabular}{lccc}
Molecule & $\kappa$ & $F$ & $B$ \\
\hline 
H$_2$ & $-0.221$ & $0.429$  & 2.28 \\
N$_2$ & $-0.115$ & 0.406 & 3.96 \\
Cl$_2$ ($\rho_\text{Cl}=1.88$~a.u.) & $+0.0349$ & 1.39 & 6.28 \\
Cl$_2$ ($\rho_\text{Cl}=2.20$~a.u.) & $-0.0187$ & 0.778 & 7.36 \\
CH$_4$ & $-0.0545$ & 0.805 & 5.24
\end{tabular}
\end{ruledtabular}
\end{table}
Finally, the figure shows the Kohn-variational calculation of Armour and Baker \cite{Armour86}, with which we obtain very good agreement.
Our predicted zero-energy value of $\Zeff$ is 11.1. This is lower than the results of existing stochastic variational calculations, which gave $\Zeff=15.7$ \cite{Zhang09,Zhang11} and 15.8 \cite{Zhang14}, and should be regarded as more accurate.
At thermal (room-temperature) energies, corresponding to the momentum $k=0.053$~a.u., we obtain $\Zeff = 10.6$. This is close to the $R$-matrix value of 10.4 \cite{Zhang11a}, but smaller than the recommended experimental room-temperature value of $\Zeff=16.0\pm0.2$ \cite{Charlton13}.

\begin{figure}
\centering
\includegraphics[width=0.4\textwidth]{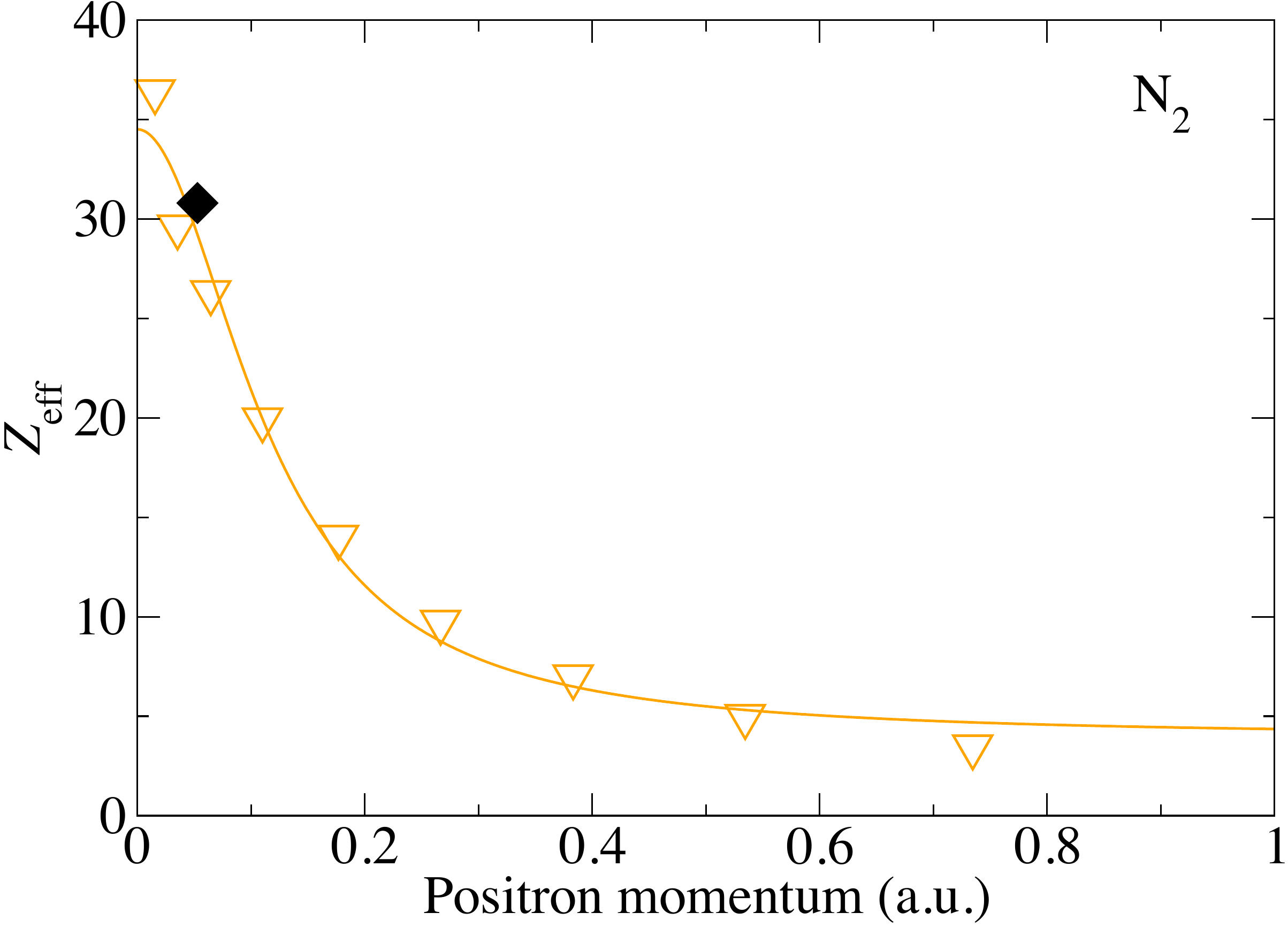} \\[0.5em]
\includegraphics[width=0.4\textwidth]{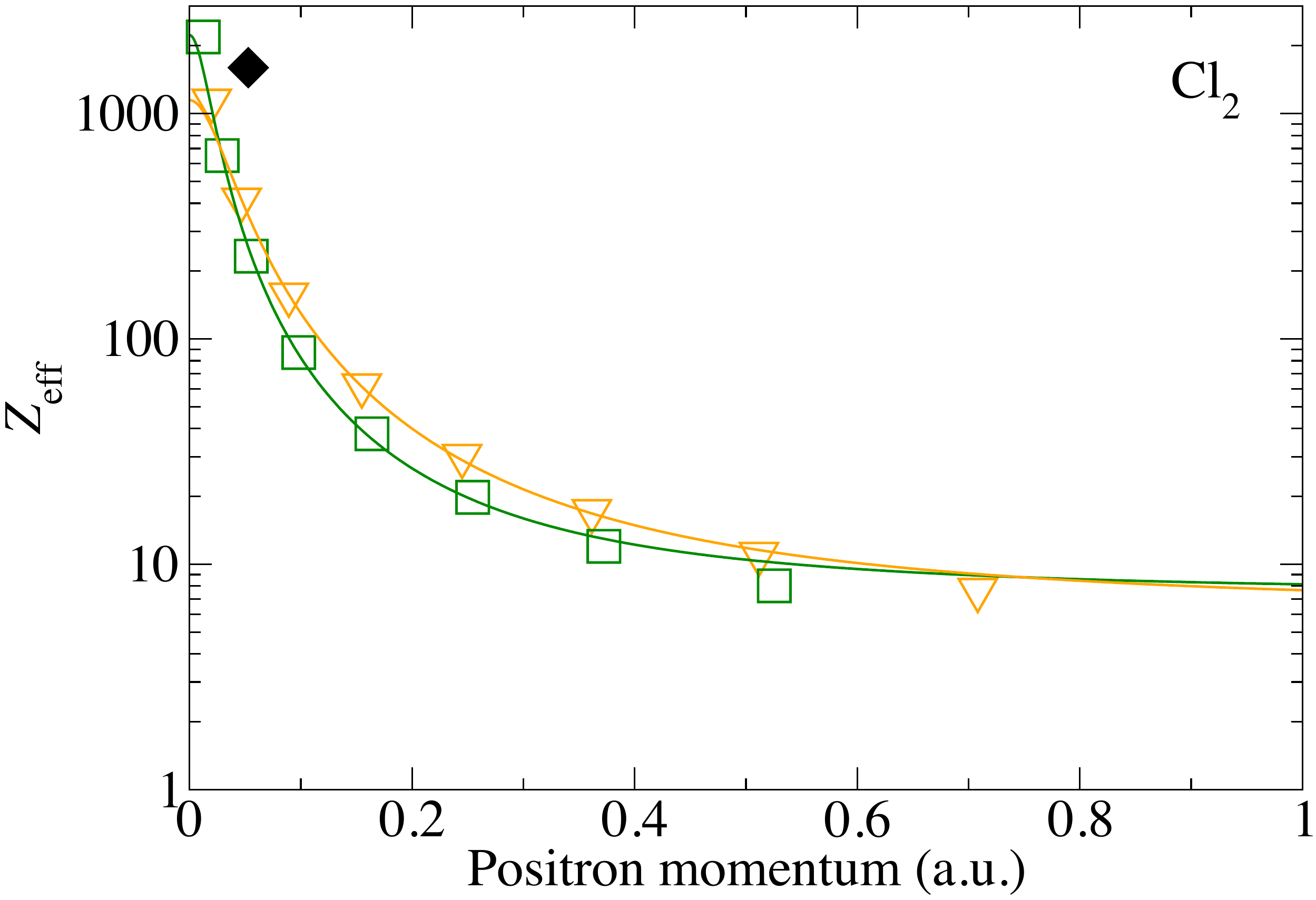} \\[0.5em]
\includegraphics[width=0.4\textwidth]{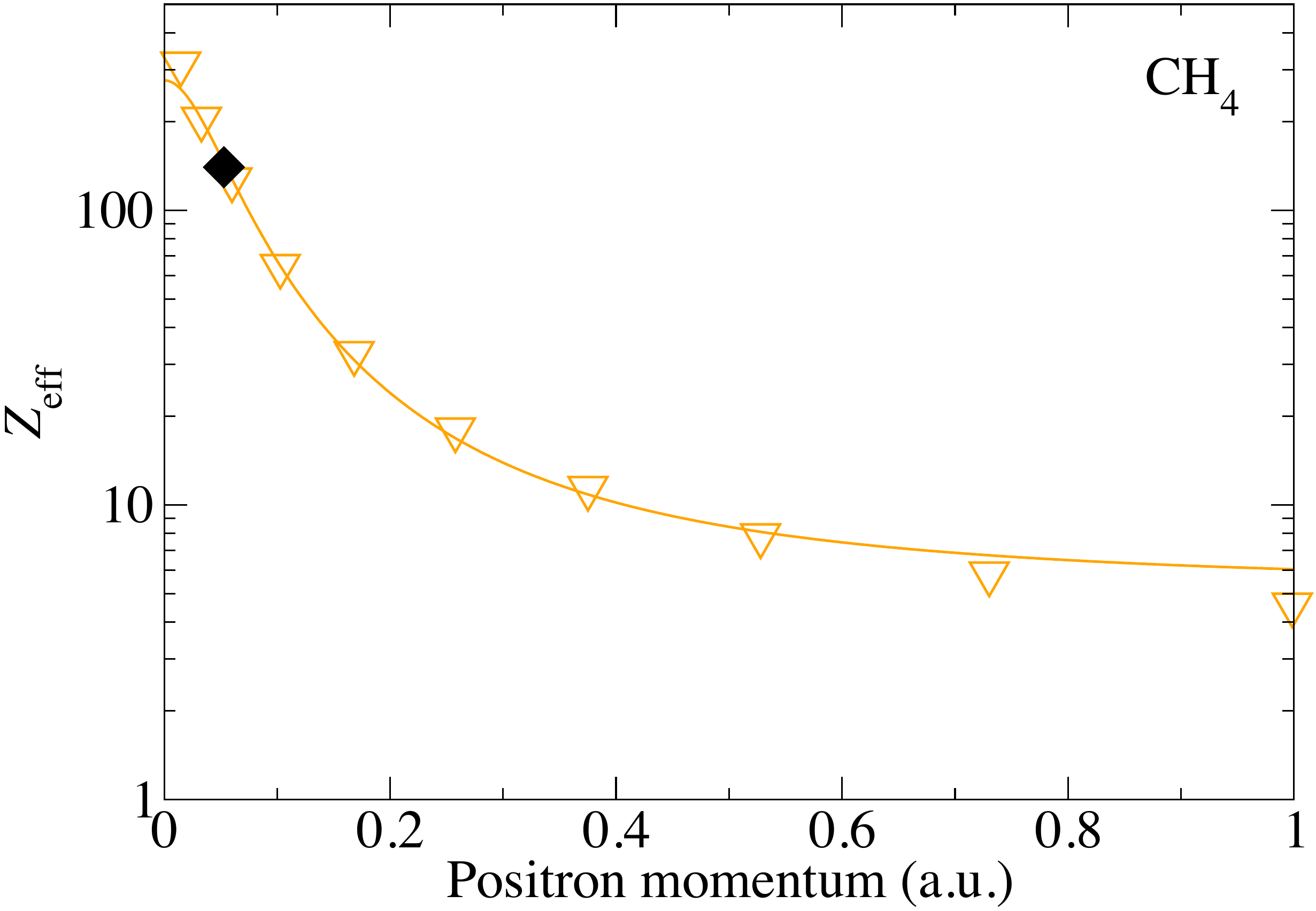}
\caption{\label{fig:Zeff_N2Cl2CH4}$\Zeff$ for N$_2$, Cl$_2$, and CH$_4$. Orange down triangles; $12s\,8p\,8d$ calculation for N$_2$ and Cl$_2$ ($\rho_\text{Cl}=1.88$~a.u.), and $12s\,8p\,8d\,/\,8s$ calculation for CH$_4$; green squares, $12s\,8p\,8d$ calculation for  Cl$_2$ ($\rho_\text{Cl}=2.20$~a.u.) (all using analytical estimate for normalization); solid  lines, fits of the form (\ref{eq:Zeff_k_dep});
black diamond, experimental room-temperature values, shown at thermal $k=0.053$~a.u. \cite{Tao65,Charlton13}.}
\end{figure}

The momentum dependence of $\Zeff$ for N$_2$, Cl$_2$, and CH$_4$ is shown in Fig.~\ref{fig:Zeff_N2Cl2CH4}.
Again, fits of the form (\ref{eq:Zeff_k_dep}) were carried out, with the first pseudostate excluded from the fit. The fitting parameters are given in Table \ref{tab:Zeff_molecules}. The values of $\kappa$ obtained from the fit can be used to verify the scattering lengths for Cl$_2$ and CH$_4$. In the presence of a weakly bound or low-lying virtual level, the value of $\kappa $ in the momentum dependence of $\Zeff$, Eq.~(\ref{eq:Zeff_k_dep}), is related to the scattering length by $a\approx1/\kappa$. Using $\kappa $ from Table~\ref{tab:Zeff_molecules}, we find $a=28.7$~a.u. or $-53.5$~a.u. for Cl$_2$ with $\rho_\text{Cl}=1.88$ or 2.20~a.u., respectively, and $a=-18.3$~a.u. for CH$_4$. The values for Cl$_2$ are close to the estimates obtained from
the binding energy (for $\rho_\text{Cl}=1.88$) and extrapolation of the phase shifts.
The value for CH$_4$ is almost within the error bars of the scattering length obtained from the phase shift.

\begin{figure}
\centering
\includegraphics[width=0.4\textwidth]{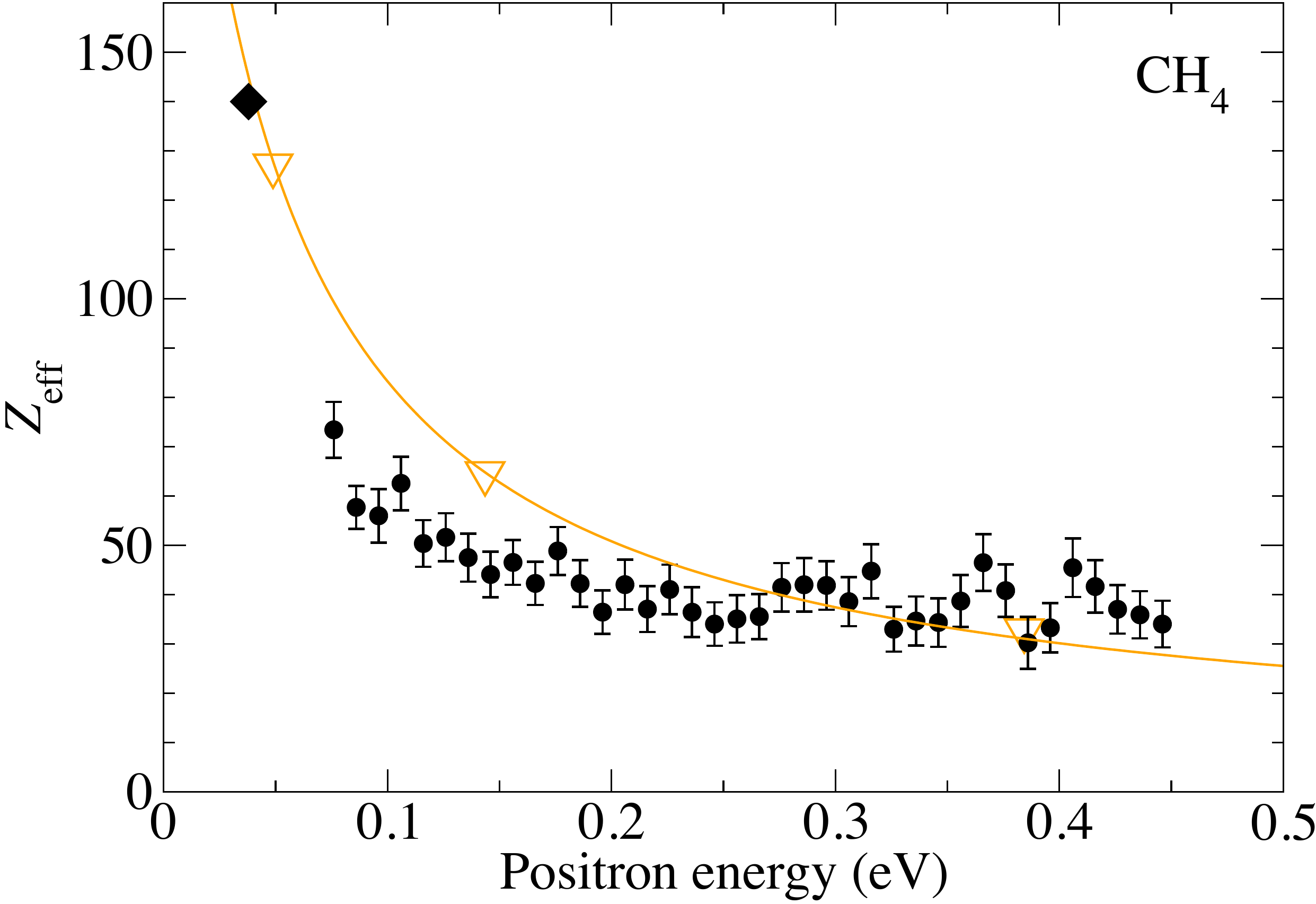}
\caption{\label{fig:Zeff_CH4_zoom}$\Zeff$ for  CH$_4$. Orange down triangles; $12s\,8p\,8d$ calculation using analytical estimate for normalization (same as in Fig.~\ref{fig:Zeff_N2Cl2CH4}); solid orange line, fit of the form (\ref{eq:Zeff_k_dep}); black circles, experimental data \cite{Barnes03,Marler04}; black diamond, experimental room-temperature value shown at thermal energy (0.0379~eV) \cite{Charlton13}.}
\end{figure}

The calculated values of $\Zeff$ at thermal momentum $k=0.053$~a.u., are 29.3, 351, 254, and 145 for N$_2$, Cl$_2$ ($\rho_\text{Cl}=1.88$~a.u.), Cl$_2$ ($\rho_\text{Cl}=2.20$~a.u.), and CH$_4$, respectively. The  experimental values, which are also shown in Fig.~\ref{fig:Zeff_N2Cl2CH4}, are $30.8\pm0.2$ for N$_2$ \cite{Charlton13}, 1600 for Cl$_2$ \cite{Tao65}, and $140.0\pm0.8$ for CH$_4$ \cite{Charlton13}.
The agreement with experiment is excellent for both N$_2$ and  CH$_4$. For CH$_4$ we also compare the calculated $\Zeff$ with energy-resolved annihilation measurements  \cite{Barnes03,Marler04}; see Fig.~\ref{fig:Zeff_CH4_zoom}.
The present calculation indicates a slightly stronger energy dependence for $\Zeff$ than seen in the experiment, but the overall agreement is very satisfactory.

For Cl$_2$, the calculated $\Zeff$ values are much smaller than the measured room-temperature value of $\Zeff=1600$ \cite{Tao65}. Unfortunately, this early measurement has never been repeated by other experimental groups, so one may query the accuracy of this large value. On the side of theory, the calculated values represent only the contribution of direct, in-flight annihilation to $\Zeff $. There is not much difference between the
$\Zeff$ obtained using $\rho_\text{Cl}=1.88$ and 2.20~a.u. However, the calculation with the smaller cutoff radius, predicts that the positron has a bound state with Cl$_2$.
In this case, the \textit{resonant} annihilation mechanism operates alongside direct annihilation \cite{Gribakin00,Gribakin10}. For a molecule with one vibrational mode, the thermally avergaged contribution of resonant annihilation to $\Zeff$ can be estimated as
\cite{GribakinNewDirections}
\al{\label{eq:Zeff_res}
\bar Z_\text{eff}^{\text{(res)}}(T) = \frac{8\pi^3 \delta_{ep}}{(2\pi k_B T)^{3/2}} \frac{e^{\epsilon _b / k_B T}}{e^{\omega / k_B T}-1} ,
}
where $T$ is the temperature, $k_B$ is the Boltzmann constant, $\delta_{ep}$ is the electron-positron contact density in the bound state, $\omega$ is the frequency of the vibrational mode of the positron-molecule complex, and $\omega >\epsilon _b$ is assumed.
Using the values of $\eb$ and $\delta_{ep}$ found earlier for the $e^+$Cl$_2$ bound state, and the vibrational frequency of Cl$_2$, $\omega=560~\text{cm}^{-1}=2.55\times10^{-3}$~a.u. \cite{CRC}, we find at room temperature $T=293$~K, 
$\bar Z _\text{eff}^{\text{(res)}}(T)=316$. Adding this to the corresponding direct contribution, gives the total value of $\Zeff = 667$, which is still significantly smaller than the measured value. In principle, the resonant annihilation contribution can be made bigger by allowing for a larger binding energy $\epsilon _b$. This will increase both the contact density $\delta _{ep}$ and the Boltzmann-type factor in Eq.~(\ref{eq:Zeff_res}). However, increasing the binding energy requires a smaller value of $\rho_\text{Cl}$, which is hard to justify physically.
In all of the present calculations of $\Zeff$, we have neglected the rotational motion of the molecule. Of course, in room-temperature measurements of $\Zeff$, the molecule can be in a variety of rotationally excited states. However, for $s$-wave positron attachment, the rotational state of the molecule does not change, and the $\Zeff$ is not expected to be noticeably affected by molecular rotations.

\section{Conclusions}

A model-potential approach has been used to study low-energy positron interactions with a range of atoms and small nonpolar molecules. The positron-target correlation potential that we use accounts for long-range polarization of the target. Short-range correlations are parametrized by a cutoff radius whose values can be specific for each type of atom within the target. These  values can be chosen to reproduce existing accurate calculations of positron binding or scattering from atomic targets, or other data, e.g., measured positron-molecule binding energies.
Positron binding energies and bound-state annihilation rates (where bound states exist), scattering phase shifts, scattering lengths, and the annihilation parameter $\Zeff$ have been calculated. The results compare very favorably with existing calculations and experimental data. In particular, we have obtained  $\Zeff $ values for N$_2$ and CH$_4$ in excellent agreement with well-established room-temperature values. For CH$_4$, our calculations confirmed the role played by the low-lying virtual state in producing enhanced $\Zeff $ at low positron energies, which was conjectured long time ago \cite{Goldanskii64}. One exception is Cl$_2$, where the present $\Zeff$ strongly underestimates the early experimental data. This discrepancy remains an open question, as it is not clear that even the presence of a weakly bound positron state and associated resonant annihilation can bridge the gap between theory and experiment.

On the technical side, the calculations for atoms confirmed the applicability of Gaussian bases to the problem of positron binding. More importantly, we have shown how to use positive-energy pseudostates to study positron scattering and direct annihilation. Our calculations also proved the validity of enhancement factors for the calculation of annihilation rates, i.e., the contact density $\delta_{ep}$ for the positron bound states and $\Zeff$ for positron scattering.

Although the theoretical description of the positron-molecule interaction is not \textit{ab initio}, it appears to capture the essential physics of the positron-molecule problem correctly. The advantage of the present technique over \textit{ab initio} methods (which have so far failed to accurately predict positron-molecule binding energies)
is that it can easily be used for large molecules, with very little computational expense.
The main source of uncertainty in the binding energies and contact densities, and in the phase shifts and $\Zeff$ at low energies, is the choice of cutoff radii for the model correlation potential. However, if the cutoff radius for each type of atom can be chosen by reference to an accurate \emph{ab initio} calculation of the positron binding energy or $s$-wave phase shift, then using these cutoff radii in molecular systems is expected to give reliable results, as we have have demonstrated, e.g., in calculating the phase shift for H$_2$. An additional source uncertainty of the present approach is that for larger, non--spherical-top molecules, the molecular polarizability tensor can be significantly anisotropic, which our correlation potential does not account for. In any case, we expect that the uncertainty in our results for positron binding or low-energy scattering should not exceed 10--20\%. The use of a parametrized formula for the annihilation enhancement factors also introduces some uncertainty in the contact densities and $\Zeff$, although the agreement of our calculated $s$-wave $\Zeff$ for H$_2$ with the Kohn-variational calculations, and of our thermal $\Zeff$ for N$_2$ and CH$_4$ with the experimental data, indicates that the formula describes the enhancement very well.

We intend to use the method to investigate positron binding to other larger molecules, in particular, large  species for which there are no existing \textit{ab initio} calculations. (Our earlier use of the method to investigate positron binding to alkane molecules was very successful in reproducing the experimental trends \cite{Swann19}.) In addition to having values of the cutoff radii for C and H atoms \cite{Swann18,Swann19}, this work has provided values for N and Cl atoms that can be used to investigate positron binding to nitriles and chlorinated hydrocarbons (although the value for Cl is more tentative), for which some experimental measurements of the binding energy already exist \cite{Young07,Danielson12}. Determining an appropriate value of the cutoff radius for an O atom would enable calculations for alcohol, aldehydes, ketones, formates, and acetates. It should also be possible to use the method to calculate $\Zeff$ for polar molecules.

In addition to computing binding energies and bound-state annihilation rates, scattering phase shifts, and $\Zeff$, for molecules that bind the positron we will calculate the annihilation $\gamma$-ray spectra, for which much of the experimental data \cite{Iwata97} remained unexplained for a long time \cite{Green12} and are only starting to be investigated now \cite{Ikabata18}.

\begin{acknowledgments}
We are grateful to C. M. Surko and J. R. Danielson for useful discussions.
This work  has been supported by the EPSRC UK, Grant No. EP/R006431/1. 
\end{acknowledgments}

\appendix

\section{\label{sec:L2expec}Expectation values of {$L^2$} operator in a Gaussian basis}

The expectation value of the squared-angular-momentum operator $L^2$ for a positron in a state with wave function $\psi(\vec{r})$, which has been expressed using a Gaussian basis [see Eqs.~(\ref{eq:pos_wfn_exp}) and (\ref{eq:gaussian_primitive})], is given by
\al{\label{eq:a1}
\langle L^2 \rangle = \sum_{B=1}^{N_a} \sum_{k=1}^{N^p_B} \sum_{A=1}^{N_a} \sum_{j=1}^{N^p_A} \big[C_{Bk}^{(p)} \big]^* C_{Aj}^{(p)}
\langle Bk \vert L^2 \vert Aj \rangle ,
}
where
\al{\label{eq:a2}
\langle Bk \vert L^2 \vert Aj \rangle = \int g_{Bk}(\vec{r}) L^2 g_{Aj}(\vec{r}) \, d\tau.
}
For brevity, we will combine the indices $A$ and $B$ that enumerate the nuclei with the corresponding indices $j$ and $k$ that enumerate the basis functions centered on each nucleus into single indices that enumerate all of the basis functions across all centers. We also drop the superscript $(p)$ from the expansion coefficients of the positron wave function. Equation (\ref{eq:a1})  becomes
\al{
\langle L^2 \rangle = \sum_{k} \sum_j C_k^* C_j \langle k \vert L^2 \vert j \rangle ,
}
where
\al{
\langle k \vert L^2 \vert j \rangle &= \int g_k(\vec{r}) L^2 g_j(\vec{r}) \, d\tau \notag\\
&=  \int_{-\infty}^\infty \int_{-\infty}^\infty \int_{-\infty}^\infty
(x-x_k)^{n^x_k} (y-y_k)^{n^y_k} (z-z_k)^{n^z_k} \notag\\
&\quad{}\times \exp\big\{{-}\zeta_k \big[(x-x_k)^2+(y-y_k)^2+(z-z_k)^2\big] \big\} \notag\\
&\quad{}\times L^2 \Big(
(x-x_j)^{n^x_j} (y-y_j)^{n^y_j} (z-z_j)^{n^z_j} \notag\\
&\quad{}\times \exp\big\{{-}\zeta_j \big[(x-x_j)^2+(y-y_j)^2+(z-z_j)^2\big] \big\} 
\Big) \notag\\
&\quad{}\times dx \, dy \, dz.
}
The standard expression for the $L^2$ operator in Cartesian coordinates is
\al{
L^2 &= 2x \pd{}{x} + 2y \pd{}{y} + 2z \pd{}{z} \notag\\
& \quad{} + 2xy \pdd{}{x}{y} + 2xz \pdd{}{x}{z} + 2yz \pdd{}{y}{z} \notag\\
& \quad{} - x^2 \pds{}{y} - x^2 \pds{}{z} - y^2 \pds{}{x} \notag \\
& \quad{} - y^2 \pds{}{z} - z^2 \pds{}{x} - z^2 \pds{}{y} ,
}
so we require expressions for the integrals $\langle k \vert 2x \, \partial / \partial x \vert j \rangle$, $\langle k \vert 2y \, \partial / \partial y \vert j \rangle$, etc.
We define the overlap integral $\langle k\vert j\rangle$ between two Gaussian basis functions, and for later convenience, we explicitly show the powers $n^x_j$, $n^y_j$, and $n^z_j$ \cite{Swann18}:
\begin{widetext}
\al{
\langle k\vert j\rangle \equiv
\langle k \vert j ,n^x_j, n^y_j, n^z_j \rangle &= \int g_k(\vec{r})  g_j(\vec{r}) \, d\tau \notag\\
&=  \int_{-\infty}^\infty \int_{-\infty}^\infty \int_{-\infty}^\infty
(x-x_k)^{n^x_k} (y-y_k)^{n^y_k} (z-z_k)^{n^z_k}  \exp\big\{{-}\zeta_k \big[(x-x_k)^2+(y-y_k)^2+(z-z_k)^2\big] \big\} \notag\\
&\quad{}\times 
(x-x_j)^{n^x_j} (y-y_j)^{n^y_j} (z-z_j)^{n^z_j}  \exp\big\{{-}\zeta_j \big[(x-x_j)^2+(y-y_j)^2+(z-z_j)^2\big] \big\} \, dx \, dy \, dz  \notag\\
&=  e^{-\lambda_{jk} \lvert \vec{r}_j - \vec{r}_k\rvert^2} \prod_{\mu=x,y,z} \sum_{s^\mu_j=0}^{n^\mu_j} \sum_{s^\mu_k=0}^{n^\mu_k}
\binom{n^\mu_j}{s^\mu_j} \binom{n^\mu_k}{s^\mu_k}
\frac12 \big[ 1+(-1)^{s^\mu_j+s^\mu_k}\big]
(\mu_{jk} - \mu_j)^{n^\mu_j-s^\mu_j} (\mu_{jk} - \mu_k)^{n^\mu_k-s^\mu_k} \notag\\
&\quad{} \times (\zeta_j + \zeta_k)^{-(1+s^\mu_j+s^\mu_k)/2} \Gamma \left( \frac{1 + s^\mu_j+s^\mu_k}{2} \right) ,
}
\end{widetext}
where
\al{
\lambda_{jk} &= \frac{\zeta_j \zeta_k}{\zeta_j + \zeta_k} , \\
\mu_{jk} &= \frac{\zeta_j \mu_j + \zeta_k \mu_k}{\zeta_j + \zeta_k} \quad (\mu= x,y,z).
}
After lengthy computation, we obtain
\al{
\left\langle k \middle\vert 2x\pd{}{x} \middle\vert j \right\rangle &= 
2 n^x_j \langle k \vert j ,n^x_j, n^y_j, n^z_j \rangle \notag\\ 
&\quad{} - 4 \zeta_j \langle k \vert j ,n^x_j+2, n^y_j, n^z_j \rangle \notag\\
&\quad{} + 2 n^x_j x_j \langle k \vert j ,n^x_j-1, n^y_j, n^z_j \rangle \notag\\
&\quad{} - 4\zeta_j x_j \langle k \vert j ,n^x_j+1, n^y_j, n^z_j \rangle ,
}
\al{
\left\langle k \middle\vert 2y\pd{}{y} \middle\vert j \right\rangle &= 
 2 n^y_j \langle k \vert j ,n^x_j, n^y_j, n^z_j \rangle \notag\\
&\quad{} - 4 \zeta_j \langle k \vert j ,n^x_j, n^y_j+2, n^z_j \rangle \notag\\
&\quad{} + 2 n^y_j y_j \langle k \vert j ,n^x_j, n^y_j-1, n^z_j \rangle \notag\\
&\quad{} - 4\zeta_j y_j \langle k \vert j ,n^x_j, n^y_j+1, n^z_j \rangle ,
}
\al{
\left\langle k \middle\vert 2z\pd{}{z} \middle\vert j \right\rangle &= 
 2 n^z_j \langle k \vert j ,n^x_j, n^y_j, n^z_j \rangle \notag\\
&\quad{} - 4 \zeta_j \langle k \vert j ,n^x_j, n^y_j, n^z_j+2 \rangle \notag\\
&\quad{} + 2 n^z_j z_j \langle k \vert j ,n^x_j, n^y_j, n^z_j-1 \rangle \notag\\
&\quad{} - 4\zeta_j z_j \langle k \vert j ,n^x_j, n^y_j, n^z_j+1 \rangle ,
}
\al{
\left\langle k \middle\vert 2xy\pdd{}{x}{y} \middle\vert j \right\rangle &= 
 2 n^x_j n^y_j \langle k \vert j ,n^x_j, n^y_j, n^z_j \rangle \notag\\
&\quad{} - 4 \zeta_j n^x_j \langle k \vert j ,n^x_j, n^y_j+2, n^z_j \rangle \notag\\
&\quad{} - 4 \zeta_j n^y_j \langle k \vert j ,n^x_j+2, n^y_j, n^z_j \rangle \notag\\
&\quad{} + 8 \zeta_j^2 \langle k \vert j ,n^x_j+2, n^y_j+2, n^z_j \rangle \notag\\
&\quad{} + 2 n^x_j n^y_j x_j \langle k \vert j ,n^x_j-1, n^y_j, n^z_j \rangle \notag\\
&\quad{} - 4 \zeta_j n^x_j x_j \langle k \vert j ,n^x_j-1, n^y_j+2, n^z_j \rangle \notag\\
&\quad{} - 4 \zeta_j n^y_j x_j \langle k \vert j ,n^x_j+1, n^y_j, n^z_j \rangle \notag\\
&\quad{} + 8 \zeta_j^2 x_j \langle k \vert j ,n^x_j+1, n^y_j+2, n^z_j \rangle \notag\\
&\quad{} + 2 n^x_j n^y_j y_j \langle k \vert j ,n^x_j, n^y_j-1, n^z_j \rangle \notag\\
&\quad{} - 4 \zeta_j n^x_j y_j \langle k \vert j ,n^x_j, n^y_j+1, n^z_j \rangle \notag\\
&\quad{} - 4 \zeta_j n^y_j y_j \langle k \vert j ,n^x_j+2, n^y_j-1, n^z_j \rangle \notag\\
&\quad{} + 8 \zeta_j^2 y_j \langle k \vert j ,n^x_j+2, n^y_j+1, n^z_j \rangle \notag\\
&\quad{} + 2 n^x_j n^y_j x_j y_j \langle k \vert j ,n^x_j-1, n^y_j-1, n^z_j \rangle \notag\\
&\quad{} - 4 \zeta_j n^x_j x_j y_j \langle k \vert j ,n^x_j-1, n^y_j+1, n^z_j \rangle \notag\\
&\quad{} - 4 \zeta_j n^y_j x_j y_j \langle k \vert j ,n^x_j+1, n^y_j-1, n^z_j \rangle \notag\\
&\quad{} + 8 \zeta_j^2 x_j y_j \langle k \vert j ,n^x_j+1, n^y_j+1, n^z_j \rangle ,
}
\al{
\left\langle k \middle\vert 2xz\pdd{}{x}{z} \middle\vert j \right\rangle &= 
 2 n^x_j n^z_j \langle k \vert j ,n^x_j, n^y_j, n^z_j \rangle \notag\\
&\quad{} - 4 \zeta_j n^x_j \langle k \vert j ,n^x_j, n^y_j, n^z_j+2 \rangle \notag\\
&\quad{} - 4 \zeta_j n^z_j \langle k \vert j ,n^x_j+2, n^y_j, n^z_j \rangle \notag\\
&\quad{} + 8 \zeta_j^2 \langle k \vert j ,n^x_j+2, n^y_j, n^z_j+2 \rangle \notag\\
&\quad{} + 2 n^x_j n^z_j x_j \langle k \vert j ,n^x_j-1, n^y_j, n^z_j \rangle \notag\\
&\quad{} - 4 \zeta_j n^x_j x_j \langle k \vert j ,n^x_j-1, n^y_j, n^z_j+2 \rangle \notag\\
&\quad{} - 4 \zeta_j n^z_j x_j \langle k \vert j ,n^x_j+1, n^y_j, n^z_j \rangle \notag\\
&\quad{} + 8 \zeta_j^2 x_j \langle k \vert j ,n^x_j+1, n^y_j, n^z_j+2 \rangle \notag\\
&\quad{} + 2 n^x_j n^z_j z_j \langle k \vert j ,n^x_j, n^y_j, n^z_j-1 \rangle \notag\\
&\quad{} - 4 \zeta_j n^x_j z_j \langle k \vert j ,n^x_j, n^y_j, n^z_j+1 \rangle \notag\\
&\quad{} - 4 \zeta_j n^z_j z_j \langle k \vert j ,n^x_j+2, n^y_j, n^z_j-1 \rangle \notag\\
&\quad{} + 8 \zeta_j^2 z_j \langle k \vert j ,n^x_j+2, n^y_j, n^z_j+1 \rangle \notag\\
&\quad{} + 2 n^x_j n^z_j x_j z_j \langle k \vert j ,n^x_j-1, n^y_j, n^z_j-1 \rangle \notag\\
&\quad{} - 4 \zeta_j n^x_j x_j z_j \langle k \vert j ,n^x_j-1, n^y_j, n^z_j+1 \rangle \notag\\
&\quad{} - 4 \zeta_j n^z_j x_j z_j \langle k \vert j ,n^x_j+1, n^y_j, n^z_j-1 \rangle \notag\\
&\quad{} + 8 \zeta_j^2 x_j z_j \langle k \vert j ,n^x_j+1, n^y_j, n^z_j+1 \rangle ,
}
\al{
\left\langle k \middle\vert 2yz\pdd{}{y}{z} \middle\vert j \right\rangle &= 
 2 n^y_j n^z_j \langle k \vert j ,n^x_j, n^y_j, n^z_j \rangle \notag\\
&\quad{} - 4 \zeta_j n^y_j \langle k \vert j ,n^x_j, n^y_j, n^z_j+2 \rangle \notag\\
&\quad{} - 4 \zeta_j n^z_j \langle k \vert j ,n^x_j, n^y_j+2, n^z_j \rangle \notag\\
&\quad{} + 8 \zeta_j^2 \langle k \vert j ,n^x_j, n^y_j+2, n^z_j+2 \rangle \notag\\
&\quad{} + 2 n^y_j n^z_j y_j \langle k \vert j ,n^x_j, n^y_j-1, n^z_j \rangle \notag\\
&\quad{} - 4 \zeta_j n^y_j y_j \langle k \vert j ,n^x_j, n^y_j-1, n^z_j+2 \rangle \notag\\
&\quad{} - 4 \zeta_j n^z_j y_j \langle k \vert j ,n^x_j, n^y_j+1, n^z_j \rangle \notag\\
&\quad{} + 8 \zeta_j^2 y_j \langle k \vert j ,n^x_j, n^y_j+1, n^z_j+2 \rangle \notag\\
&\quad{} + 2 n^y_j n^z_j z_j \langle k \vert j ,n^x_j, n^y_j, n^z_j-1 \rangle \notag\\
&\quad{} - 4 \zeta_j n^y_j z_j \langle k \vert j ,n^x_j, n^y_j, n^z_j+1 \rangle \notag\\
&\quad{} - 4 \zeta_j n^z_j z_j \langle k \vert j ,n^x_j, n^y_j+2, n^z_j-1 \rangle \notag\\
&\quad{} + 8 \zeta_j^2 z_j \langle k \vert j ,n^x_j, n^y_j+2, n^z_j+1 \rangle \notag\\
&\quad{} + 2 n^y_j n^z_j y_j z_j \langle k \vert j ,n^x_j, n^y_j-1, n^z_j-1 \rangle \notag\\
&\quad{} - 4 \zeta_j n^y_j y_j z_j \langle k \vert j ,n^x_j, n^y_j-1, n^z_j+1 \rangle \notag\\
&\quad{} - 4 \zeta_j n^z_j y_j z_j \langle k \vert j ,n^x_j, n^y_j+1, n^z_j-1 \rangle \notag\\
&\quad{} + 8 \zeta_j^2 y_j z_j \langle k \vert j ,n^x_j, n^y_j+1, n^z_j+1 \rangle ,
}
\al{
\left\langle k \middle\vert -x^2\pds{}{y} \middle\vert j \right\rangle &=
- n^y_j (n^y_j-1) \langle k \vert j ,n^x_j+2, n^y_j-2, n^z_j \rangle \notag\\
&\quad{} + 2 \zeta_j (2 n^y_j+1 ) \langle k \vert j ,n^x_j+2, n^y_j, n^z_j \rangle \notag\\
&\quad{} - 4 \zeta_j^2  \langle k \vert j ,n^x_j+2, n^y_j+2, n^z_j \rangle \notag\\
&\quad{} - 2 n^y_j (n^y_j - 1) x_j  \langle k \vert j ,n^x_j+1, n^y_j-2, n^z_j \rangle \notag\\
&\quad{} + 4 \zeta_j (2 n^y_j+1 ) x_j  \langle k \vert j ,n^x_j+1, n^y_j, n^z_j \rangle \notag\\
&\quad{} - 8 \zeta_j^2 x_j  \langle k \vert j ,n^x_j+1, n^y_j+2, n^z_j \rangle \notag\\
&\quad{} - n^y_j (n^y_j-1) x_j^2 \langle k \vert j ,n^x_j, n^y_j-2, n^z_j \rangle \notag\\
&\quad{} + 2 \zeta_j (2 n^y_j+1 ) x_j^2 \langle k \vert j ,n^x_j, n^y_j, n^z_j \rangle \notag\\
&\quad{} - 4 \zeta_j^2 x_j^2 \langle k \vert j ,n^x_j, n^y_j+2, n^z_j \rangle ,
}
\al{
\left\langle k \middle\vert -x^2\pds{}{z} \middle\vert j \right\rangle &=
- n^z_j (n^z_j-1) \langle k \vert j ,n^x_j+2, n^y_j, n^z_j-2 \rangle \notag\\
&\quad{} + 2 \zeta_j (2 n^z_j+1 ) \langle k \vert j ,n^x_j+2, n^y_j, n^z_j \rangle \notag\\
&\quad{} - 4 \zeta_j^2  \langle k \vert j ,n^x_j+2, n^y_j, n^z_j+2 \rangle \notag\\
&\quad{} - 2 n^z_j (n^z_j - 1) x_j  \langle k \vert j ,n^x_j+1, n^y_j, n^z_j-2 \rangle \notag\\
&\quad{} + 4 \zeta_j (2 n^z_j+1 ) x_j  \langle k \vert j ,n^x_j+1, n^y_j, n^z_j \rangle \notag\\
&\quad{} - 8 \zeta_j^2 x_j  \langle k \vert j ,n^x_j+1, n^y_j, n^z_j+2 \rangle \notag\\
&\quad{} - n^z_j (n^z_j-1) x_j^2 \langle k \vert j ,n^x_j, n^y_j, n^z_j-2 \rangle \notag\\
&\quad{} + 2 \zeta_j (2 n^z_j+1 ) x_j^2 \langle k \vert j ,n^x_j, n^y_j, n^z_j \rangle \notag\\
&\quad{} - 4 \zeta_j^2 x_j^2 \langle k \vert j ,n^x_j, n^y_j, n^z_j+2 \rangle ,
}
\al{
\left\langle k \middle\vert -y^2\pds{}{x} \middle\vert j \right\rangle &=
- n^x_j (n^x_j-1) \langle k \vert j ,n^x_j-2, n^y_j+2, n^z_j \rangle \notag\\
&\quad{} + 2 \zeta_j (2 n^x_j+1 ) \langle k \vert j ,n^x_j, n^y_j+2, n^z_j \rangle \notag\\
&\quad{} - 4 \zeta_j^2  \langle k \vert j ,n^x_j+2, n^y_j+2, n^z_j \rangle \notag\\
&\quad{} - 2 n^x_j (n^x_j - 1) y_j  \langle k \vert j ,n^x_j-2, n^y_j+1, n^z_j \rangle \notag\\
&\quad{} + 4 \zeta_j (2 n^x_j+1 ) y_j  \langle k \vert j ,n^x_j, n^y_j+1, n^z_j \rangle \notag\\
&\quad{} - 8 \zeta_j^2 y_j  \langle k \vert j ,n^x_j+2, n^y_j+1, n^z_j \rangle \notag\\
&\quad{} - n^x_j (n^x_j-1) y_j^2 \langle k \vert j ,n^x_j-2, n^y_j, n^z_j \rangle \notag\\
&\quad{} + 2 \zeta_j (2 n^x_j+1 ) y_j^2 \langle k \vert j ,n^x_j, n^y_j, n^z_j \rangle \notag\\
&\quad{} - 4 \zeta_j^2 y_j^2 \langle k \vert j ,n^x_j+2, n^y_j, n^z_j \rangle ,
}
\al{
\left\langle k \middle\vert -y^2\pds{}{z} \middle\vert j \right\rangle &=
- n^z_j (n^z_j-1) \langle k \vert j ,n^x_j, n^y_j+2, n^z_j-2 \rangle \notag\\
&\quad{} + 2 \zeta_j (2 n^z_j+1 ) \langle k \vert j ,n^x_j, n^y_j+2, n^z_j \rangle \notag\\
&\quad{} - 4 \zeta_j^2  \langle k \vert j ,n^x_j, n^y_j+2, n^z_j+2 \rangle \notag\\
&\quad{} - 2 n^z_j (n^z_j - 1) y_j  \langle k \vert j ,n^x_j, n^y_j+1, n^z_j-2 \rangle \notag\\
&\quad{} + 4 \zeta_j (2 n^z_j+1 ) y_j  \langle k \vert j ,n^x_j, n^y_j+1, n^z_j \rangle \notag\\
&\quad{} - 8 \zeta_j^2 y_j  \langle k \vert j ,n^x_j, n^y_j+1, n^z_j+2 \rangle \notag\\
&\quad{} - n^z_j (n^z_j-1) y_j^2 \langle k \vert j ,n^x_j, n^y_j, n^z_j-2 \rangle \notag\\
&\quad{} + 2 \zeta_j (2 n^z_j+1 ) y_j^2 \langle k \vert j ,n^x_j, n^y_j, n^z_j \rangle \notag\\
&\quad{} - 4 \zeta_j^2 y_j^2 \langle k \vert j ,n^x_j, n^y_j, n^z_j+2 \rangle ,
}
\al{
\left\langle k \middle\vert -z^2\pds{}{x} \middle\vert j \right\rangle &=
- n^x_j (n^x_j-1) \langle k \vert j ,n^x_j-2, n^y_j, n^z_j+2 \rangle \notag\\
&\quad{} + 2 \zeta_j (2 n^x_j+1 ) \langle k \vert j ,n^x_j, n^y_j, n^z_j+2 \rangle \notag\\
&\quad{} - 4 \zeta_j^2  \langle k \vert j ,n^x_j+2, n^y_j, n^z_j+2 \rangle \notag\\
&\quad{} - 2 n^x_j (n^x_j - 1) z_j  \langle k \vert j ,n^x_j-2, n^y_j, n^z_j+1 \rangle \notag\\
&\quad{} + 4 \zeta_j (2 n^x_j+1 ) z_j  \langle k \vert j ,n^x_j, n^y_j, n^z_j+1 \rangle \notag\\
&\quad{} - 8 \zeta_j^2 z_j  \langle k \vert j ,n^x_j+2, n^y_j, n^z_j+1 \rangle \notag\\
&\quad{} - n^x_j (n^x_j-1) z_j^2 \langle k \vert j ,n^x_j-2, n^y_j, n^z_j \rangle \notag\\
&\quad{} + 2 \zeta_j (2 n^x_j+1 ) z_j^2 \langle k \vert j ,n^x_j, n^y_j, n^z_j \rangle \notag\\
&\quad{} - 4 \zeta_j^2 z_j^2 \langle k \vert j ,n^x_j+2, n^y_j, n^z_j \rangle ,
}
\al{
\left\langle k \middle\vert -z^2\pds{}{y} \middle\vert j \right\rangle &=
- n^y_j (n^y_j-1) \langle k \vert j ,n^x_j, n^y_j-2, n^z_j+2 \rangle \notag\\
&\quad{} + 2 \zeta_j (2 n^y_j+1 ) \langle k \vert j ,n^x_j, n^y_j, n^z_j+2 \rangle \notag\\
&\quad{} - 4 \zeta_j^2  \langle k \vert j ,n^x_j, n^y_j+2, n^z_j+2 \rangle \notag\\
&\quad{} - 2 n^y_j (n^y_j - 1) z_j  \langle k \vert j ,n^x_j, n^y_j-2, n^z_j+1 \rangle \notag\\
&\quad{} + 4 \zeta_j (2 n^y_j+1 ) z_j  \langle k \vert j ,n^x_j, n^y_j, n^z_j+1 \rangle \notag\\
&\quad{} - 8 \zeta_j^2 z_j  \langle k \vert j ,n^x_j, n^y_j+2, n^z_j+1 \rangle \notag\\
&\quad{} - n^y_j (n^y_j-1) z_j^2 \langle k \vert j ,n^x_j, n^y_j-2, n^z_j \rangle \notag\\
&\quad{} + 2 \zeta_j (2 n^y_j+1 ) z_j^2 \langle k \vert j ,n^x_j, n^y_j, n^z_j \rangle \notag\\
&\quad{} - 4 \zeta_j^2 z_j^2 \langle k \vert j ,n^x_j, n^y_j+2, n^z_j \rangle .
}

\section{\label{sec:spherical_averaging}Spherical averaging of positron wave function in a Gaussian basis}

The spherically averaged positron wave function is given by [see Eqs.~(\ref{eq:pos_wfn_exp}) and (\ref{eq:gaussian_primitive})]
\al{
\overline\psi(r) = \int \psi(\vec{r}) \frac{d\Omega}{4\pi} = \frac{1}{4\pi} \sum_j C_j I_j(r) ,
}
where, as in Appendix \ref{sec:L2expec}, we have combined the index $A$  that enumerates the nuclei with the  index $j$  that enumerates the basis functions centered on each nucleus into a single index that enumerates all of the basis functions across all centers, and we have dropped the superscript $(p)$ from the expansion coefficients of the positron wave function. The origin of coordinates is chosen to be at the position of the molecule's center of mass. The function $I_j$ is simply the integral of basis function $j$ over the solid angle:
\al{
I_j(r) = \int g_j(\vec{r}) \, d\Omega .
}
To find an expression for $I_j$, we use spherical polar coordinates $(r,\theta,\phi)$, which gives
\begin{widetext}
\al{\label{eq:b3}
I_j(r) &= \int_0^{2\pi} \!\! \int_0^\pi (r\sin\theta\cos\phi - x_j)^{n^x_j} (r\sin\theta\sin\phi-y_j)^{n^y_j} (r\cos\theta-z_j)^{n^z_j} \notag\\
&\quad{} \times \exp\big\{ {-}\zeta_j \big[ (r\sin\theta\cos\phi - x_j)^2 + (r\sin\theta\sin\phi-y_j)^2 + (r\cos\theta-z_j)^2 \big] \big\} \sin\theta \, d\theta \, d\phi .
}
\end{widetext}
We consider two cases: first, where only $s$-type basis functions are used, and second, where basis functions of general angular momenta are used.

\subsection{$s$-type functions only}

If only $s$-type basis functions are used, then $n^x_j=n^y_j=n^z_j=0$ for all $j$. Since an $s$-type basis function is a function only of  the distance from its center and not on the direction from its center (i.e., $g_j(\vec{r})\propto e^{-\zeta_j \lvert \vec{r}-\vec{r}_j\rvert^2}$), we are free to rotate the coordinate axes so that the center is on the $z$ axis, whence $x_j$ and $y_j$ become 0 and $z_j$ becomes $r_j$. Assuming $r_j>0$,  Eq.~(\ref{eq:b3}) becomes
\al{
I_j(r) &= e^{-\zeta_j(r^2+r_j^2)} 
\int_0^{2\pi} \!\! \int_0^\pi e^{2\zeta_j r_j r\cos\theta}\sin\theta\,d\theta\,d\phi \notag\\
&= 2\pi e^{-\zeta_j(r^2+r_j^2)}  \frac{\sinh(2\zeta_jr_jr)}{\zeta_jr_jr}.
}

If, in fact, $r_j=0$ (i.e., the basis function is centered on the origin), then Eq.~(\ref{eq:b3}) becomes
\al{
I_j(r) &= e^{-\zeta_j r^2} 
\int_0^{2\pi} \!\! \int_0^\pi \sin\theta\,d\theta\,d\phi \notag\\
&= 4\pi e^{-\zeta_j r^2} .
}

\subsection{Functions with general angular momenta}

For basis functions of general angular momenta, assuming $r_j>0$, we use the binomial theorem on the algebraic factors  in Eq.~(\ref{eq:b3}) to obtain
\begin{widetext}
\al{\label{eq:b6}
I_j(r)
&= e^{-\zeta_j (r^2 + r_j^2)}
\sum_{s^x_j=0}^{n^x_j} \sum_{s^y_j=0}^{n^y_j} \sum_{s^z_j=0}^{n^z_j}
\binom{n^x_j}{s^x_j} \binom{n^y_j}{s^y_j} \binom{n^z_j}{s^z_j}
(-x_j)^{n^x_j - s^x_j} (-y_j)^{n^y_j - s^y_j} (-z_j)^{n^z_j - s^z_j}
r^{s^x_j + s^y_j + s^z_j} \notag\\
&\quad{}\times \int_0^{2\pi} \!\! \int_0^\pi 
\cos^{s^z_j} \theta \sin^{s^x_j + s^y_j + 1}\theta \cos^{s^x_j}\phi \sin^{s^y_j}\phi
\exp \big[ 2\zeta_j r (x_j \sin\theta\cos\phi + y_j\sin\theta\sin\phi + z_j \cos\theta) \big] \, d\theta \, d\phi .
}
\end{widetext}
The integration can be carried out analytically if we restrict our interest to linear molecules. Doing this, and assuming that all nuclei (i.e., basis function centers) are positioned on the $z$ axis (so that $x_j=y_j=0$ and $r_j=\lvert z_j \rvert$ for all $j$), Eq.~(\ref{eq:b6}) simplifies to
\al{\label{eq:b7}
I_j(r) &= e^{-\zeta_j (r^2 + r_j^2)} 
\int_0^{2\pi} \cos^{n^x_j} \phi \sin^{n^y_j} \phi \, d\phi \notag\\
&\quad{}\times \sum_{s^z_j=0}^{n^z_j} \binom{n^z_j}{s^z_j} (-z_j)^{n^z_j - s^z_j} r^{n^x_j+n^y_j+s^z_j} \notag\\
&\quad{}\times \int_0^\pi \cos^{s^z_j}\theta \sin^{n^x_j+n^y_j+1}\theta\, e^{2\zeta_j z_j r\cos\theta} \, d\theta .
}
The values of the azimuthal and polar integrals depend on the parity of $n^x_j$, $n^y_j$, and $s^z_j$.

The azimuthal integral is 
\al{\label{eq:b5}
I_j^{\text{(az)}} \equiv \int_0^{2\pi} \cos^{n^x_j} \phi \sin^{n^y_j} \phi \, d\phi .
}
By splitting the domain of integration into two subintervals, $0\leq\phi\leq\pi$ and $\pi\leq\phi\leq2\pi$, and subsequently making the substitution $u=\cos\phi$ on each subinterval, we obtain
\al{
I_j^{\text{(av)}} = \big[ 1+(-1)^{n^y_j} \big] \int_{-1}^1 u^{n^x_j} (1-u^2)^{(n^y_j-1)/2} \, du .
}
Then, splitting the new domain of integration into two subintervals, $-1\leq u \leq 0$ and $0 \leq u \leq 1$,  making the substitutions $u=-\sqrt{t}$ on $-1\leq u \leq 0$ and $u=\sqrt{t}$ on $0 \leq u \leq 1$, and using the definition of the beta function,
\al{
\mathrm{B}(\alpha,\beta) = \mathrm{B}(\beta,\alpha) = \int_0^1 t^{\alpha-1} (1-t)^{\beta-1} \, dt 
}
(where $\operatorname{Re}\alpha>0$, $\operatorname{Re}\beta>0$), we obtain
\al{
I_j^{\text{(az)}} = \frac12 \big[ 1 + (-1)^{n^x_j} + (-1)^{n^y_j} + (-1)^{n^x_j+n^y_j}\big] \mathrm{B}\left( \frac{1+n^x_j}{2},\frac{1+n^y_j}{2}\right) ,
}
which gives
\al{
I_j^{\text{(az)}} = 2 \mathrm{B}\left( \frac{1+n^x_j}{2},\frac{1+n^y_j}{2}\right)
}
if $n^x_j$ and $n^y_j$ are both even, and $I_j^{\text{(az)}}=0$ otherwise.

The polar integral is solved by making the substitution $\xi=\cos\theta$:
\al{
I_j^{\text{(pol)}}(s^z_j,r) &\equiv
\int_0^\pi \cos^{s^z_j}\theta \sin^{n^x_j+n^y_j+1}\theta\, e^{2\zeta_j z_j r\cos\theta} \, d\theta ,\notag\\
&=
\int_{-1}^1 \xi^{s^z_j} (1-\xi^2)^{(n^x_j+n^y_j)/2} e^{2\zeta_j z_j r \xi} \, d\xi .
}
Splitting the domain of integration into two subintervals, $-1\leq\xi\leq0$ and $0\leq\xi\leq1$, and using the identity \cite{Gradshteyn}
\begin{widetext}
\al{
\int_0^u \xi^{2\nu-1} (u^2-\xi^2)^{\rho-1} e^{\mu\xi} \, d\xi &= 
\frac12 \mathrm{B}(\nu,\rho) u^{2\nu+2\rho-2}{}_1F_2 \left( \nu;\frac12,\nu+\rho;\frac{\mu^2 u^2}{4}\right)
+
\frac{\mu}{2} \mathrm{B}\left( \nu+\frac12,\rho\right) u^{2\nu+2\rho-1} {}_1F_2 \left( \nu+\frac12;\frac32,\nu+\rho+\frac12;\frac{\mu^2 u^2}{4}\right)
}
\end{widetext}
(which is valid for $\operatorname{Re}\rho>0$, $\operatorname{Re}\nu>0$),
we obtain
\al{\label{eq:b8}
I_j^{\text{(pol)}}(s^z_j,r)
&=
 \mathrm{B} \left(  \dfrac{2+n^x_j+n^y_j}{2} , \dfrac{1+s^z_j}{2} \right) \notag\\
&\quad{}\times {}_1F_2 \left( \frac{1+s^z_j}{2} ; \frac12 , \frac{3 + n^x_j + n^y_j + s^z_j}{2} ; \zeta_j^2 z_j^2 r^2 \right) 
} 
if $s^z_j$ is even, or
\al{\label{eq:b9}
I_j^{\text{(pol)}}(s^z_j,r)
&=
 2\zeta_j z_j r
\mathrm{B} \left( \frac{2+n^x_j+n^y_j}{2} , \frac{2+s^z_j}{2} \right) \notag\\
&\quad{}\times {}_1F_2 \left( \frac{2+s^z_j}{2} ; \frac32 , \frac{4 +n^x_j+n^y_j+s^z_j }{2} ; \zeta_j^2 z_j^2 r^2 \right)  
}
if $s^z_j$ is odd. Here, ${}_1F_2(\alpha;\beta,\gamma;\delta)$ is a generalized hypergeometric function.
The generalized hypergeometric functions that appear in Eqs.~(\ref{eq:b8}) and (\ref{eq:b9}) can, in fact, be written  as combinations of hyperbolic sines and cosines of $2\zeta_j z_j r$, i.e., we can write
\al{
I_j^{\text{(pol)}}(s^z_j,r) = A_j \sinh \rho_j + B_j \cosh \rho_j ,
}
where $\rho_j=2\zeta_j z_j r$.
In this work, we have used only $s$-, $p$-, and $d$-type basis functions; therefore, $n^x_j$, $n^y_j$, and $s^z_j$ are all integers between 0 and 2. 
Table \ref{tab:polar_integrals} shows the coefficients $A_j$ and $B_j$ for $n^x_j+n^y_j=0$, 2, 4 and $s^z_j=0$, 1, 2 (we only need to consider even values of $n^x_j + n^y_j$ since this is a requirement for the azimuthal integral  to be nonzero).
\begin{table}
\caption{\label{tab:polar_integrals}Coefficients $A_j$ and $B_j$ for the polar integral $I_j^{\text{(pol)}}(s^z_j,r)$, with $\rho_j=2\zeta_j z_j r$.}
\begin{ruledtabular}
\begin{tabular}{cccc}
$n^x_j+n^y_j$ & $s^z_j$ & $A_j$ & $B_j$ \\
\hline
0 & 0 & $2/\rho_j$ & 0 \\
  & 1 & $-2/\rho_j^2$ & $2/\rho_j$  \\
  & 2 & $2(2+\rho_j^2)/\rho_j^3$ & $-4/\rho_j^2$ \\
2 & 0 &  $-4/\rho_j^3$ & $4/\rho_j^2$ \\
   & 1 & $4(3+\rho_j^2)/\rho_j^4$ & $-12/\rho_j^3$ \\
   & 2 & $-4(12+5\rho_j^2)/\rho_j^5$ & $4(12+\rho_j^2)/\rho_j^4$ \\
4 & 0 & $16(3+\rho_j^2)/\rho_j^5$ & $-48/\rho_j^4$ \\   
  & 1 & $-48(5+2\rho_j^2)/\rho_j^6$ & $16(15+\rho_j^2)/\rho_j^5$ \\
  & 2 & $16(90+39\rho_j^2+\rho_j^4)/\rho_j^7$ & $-144(10+\rho_j^2)/\rho_j^6$
\end{tabular}
\end{ruledtabular}
\end{table}

If, in fact, $z_j=r_j=0$ (i.e., the basis function is centered on the origin), then Eq.~(\ref{eq:b3}) becomes
\al{
I_j(r) &=  r^{n^x_j+n^y_j+n^z_j} e^{-\zeta_j r^2}
\int_0^{\pi} \cos^{n^x_j}\phi \sin^{n^y_j}\phi \, d\phi \notag\\
&\quad{}\times\int_0^{2\pi} \cos^{n^z_j}\theta \sin^{n^x_j+n^y_j+1} \theta \, d\theta .
}
This gives
\al{
I_j(r) &= 2 r^{n^x_j+n^y_j+n^z_j} e^{-\zeta_j r^2} \mathrm{B}\left( \frac{1+n^x_j}{2},\frac{1+n^y_j}{2}\right) \notag\\
&\quad{}\times \mathrm{B}\left( \frac{2+n^x_j+n^y_j}{2} , \frac{1+n^z_j}{2}\right)
}
if $n^x_j$, $n^y_j$, and $n^z_j$ are all even, and $I_j(r)=0$ otherwise.

\bibliography{small_molecules_bib}

\end{document}